# Structures and Electronic States of Nickel Rich Oxides for Lithium Ion Batteries

**Saleem Yousuf[a], Md Maruf Mridha[a], and Rita Magri[a,b,c,*]**

[a]*Dipartimento di Scienze Fisiche, Informatiche e Matematiche, Università di Modena e Reggio Emilia, Via Campi 213/A, 41125 Modena, Italy*

[b]*Centro S3, Istituto Nanoscienze-Consiglio Nazionale Delle Ricerche (CNR-NANO), Via Campi 213/A, 41125 Modena, Italy*

[c]*Centro Interdipartimentale di Ricerca e per I Servizi Nel Settore Della Produzione, Stoccaggio Ed Utilizzo Dell'Idrogeno H2-MO.RE., Via Università 4, 41121, Modena, Italy*

[*]Corresponding author, *E-mail address*: rita.magri@unimore.it (R. Magri)

Keywords: Li-ion batteries, cathodes, LNO, LNMO, DFT, Layered Nickel Oxides

**Abstract**

A new superstructure of layered pristine LiNiO$_2$ (LNO) was obtained optimizing a large supercell of the $R\bar{3}m$ space group, the one observed experimentally by XRD, and relaxing both cell parameters and internal positions. The crystal structure shows size and charge disproportionation of the NiO$_6$ octahedra instead of the Jahn-Teller distortion. The decrease of the internal energy obtained with the structural optimization of the supercell relative to the same structure in its primitive unit cell is much larger than the one obtained by relaxing similarly dimensioned supercells of monoclinic symmetry relative to their primitive unit cells, although the monoclinic phase remains more stable. The Ni-O bond length distribution of the new structure agree well with the experiments. Our results show that the choice of the simulation cell is important for determining the energetics of this class of oxide materials, proposed for cathodes in lithium ion batteries (LIBs).

We used this new structure as a template for the study of the structural and electronic changes induced by the delithiation and Mn for Ni cation substitution, originating the solid solutions LiNi$_y$Mn$_{(1-y)}$O$_2$ (LNMO). Our results, surprisingly, agree well with the existing experiments and explain observed trends better than previous studies.

1. INTRODUCTION

The transition to a green economy and the environment protection needs the development of new Li-ion batteries with a higher energy density, faster to recharge, and safer. Layered oxides have been the materials of election for cathodes in Li ion batteries. Commercial Li ion batteries cathodes are made of $LiCoO_2$ (LCO), and contain cobalt, a rare and toxic element. The theoretical capacity of LCO is 270 mAhg$^{-1}$ [1] which reduces to an actual reversible charge capacity of about 145 mAhg$^{-1}$ [2] because of thermal instability issues. As a consequence, LCO needs to be substituted for large power applications. In these last years a renewed interest for the isostructural $LiNiO_2$ (LNO) compound, based on the alternative Nickel transition metal, has emerged [3]. This layered material is attractive since it has a lower cost for the same theoretical energy density of LCO. However, despite much research efforts being devoted to this material, it has not been yet commercialized, because of many drawbacks.

First, despite much care during synthesis, the material turns out to be always nonstoichiometric with Ni ions occupying Li sites so that the compound is actually $Li_{1-x}Ni_{1+x}O_2$ with $x$ ranging typically between the 2 and 10% [4]. The off-stoichiometry is thought to be due to the similarity of the $Ni^{2+}$ and $Li^+$ ionic radii (r($Ni^{2+}$) = 0.69 Å, r($Li^+$) = 0.76 Å). LNO stabilizes in the rhombohedral $R\bar{3}m$ Bravais lattice [5] despite a number of first-principles calculations have found monoclinic crystal structures to be the ground states for the layered systems [6, 7].

Other instabilities arise during the battery charge-discharge cycles, in the form of phase transitions, in which Li-Ni cation mixing leads to the formation of a spinel ($LiNi_2O_4$) phase and/or rock-salt-type phases [8, 9]. Other forms of instabilities include surface reactivity affecting the electrode-electrolyte interface with formation of NiO [10, 11]. Another critical factor governing the degradation of LNO is the multistep phase transformation process occurring on Li insertion/extraction. Four different phases have been described in the literature, namely the hexagonal H1, the monoclinic M, the hexagonal H2, and the hexagonal H3 phases [3, 4, 12, 13, 14], where the H2–H3 transformation at high state of charge (4.15–4.25 V) has the strongest negative effect on the intrinsic stability of LNO, as it is accompanied by a sudden collapse of the structure along the crystallographic c-axis. [12].

Despite the large number of studies, the structure of LNO at full Li content is still under considerable debate. The structure determined by the experiments is rhombohedral, of $R\bar{3}m$ symmetry, with an O3 stacking, where the oxygen atoms occupy the 6c sites, and lithium and nickel occupy the octahedral 3a and 3b sites [15]. In this structure the Ni cations are supposed to have $Ni^{3+}$ oxidations states in a low spin configuration, with a single electron in the $e_g$ orbitals. This electronic configuration is subjected to a Jahn-Teller (JT) distortion of the $NiO_6$ octahedrons which have been measured through EXAFS and neutron diffraction [16, 17]. The measurements supports the JT distortion: four bond lengths grouped as long bonds (2.04 Å and 2.06 Å with an average length of 2.05 Å) and short bonds (1.90 Å and 1.96 Å with an average length of 1.93 Å) suggestive of the 2:1 ratio of short-to-long bonds expected for JT distortion. However, the measurements did not indicate the existence of long range order. Indeed, a collective distortion has not been observed for LNO, as it is the case, instead of LiNaO$_2$, so it has often been assumed that in LNO the JT orbitals are randomly oriented. Experiments have later excluded the random arrangement, pointing, instead, to some kind of local ordering of a trimer kind with the three $3d_{z^2-r^2/3}$ orbitals point towards the shared oxygen site [16, 18]. First-principles calculations have confirmed that JT distorted configurations of NiO$_6$ octahedrons in LNO are more stable than the $R\bar{3}m$ undistorted structure. Chen et al. have compared the energies of differently oriented JT distorted phases finding that the structure of P2$_1$/c space group with JT distorted octahedra in a zig-zag arrangement is more stable than the C2/m structure with collinear arrangement [7]. A similar result was obtained by Das et al. [6] and Zhang et al. [19]. However, the zig-zag arrangement of the octahedra has been excluded by the experiments, finding always the rhombohedral $R\bar{3}m$ symmetry for the complete lithiated phase. Nowadays, a non-cooperative and dynamic JT effect is generally accepted for LNO [3, 20], although disproportionation based on Ni-O bond lengths has also been argued [21]. From the measured Ni-O bond lengths and by comparing them with other nickelates, a competition between charge ordering and orbital ordering for the ground state is expected for LNO.

With this paper we intend first to revisit the problem of the LNO structure using highly accurate first-principles calculations. All the simulations of the recent years studying doping and defects in LNO invariably assume for the pristine LNO model structures containing only Ni$^{3+}$ cations. In this paper we show that the structural optimization of a large supercell of $R\bar{3}m$ space group for layered LNO leads to an ordered crystal structure belonging again to the $R\bar{3}m$ space group but presenting structural and charge disproportionation. The structural optimization leads to a large energy gain much larger than the energy gain obtained for the *C2/m* monoclinic space group in a similarly large supercell. Still. the monoclinic crystal structure has a lower energy than the rhombohedral one but the difference between their energies becomes much smaller. This result is obtained for pristine LNO without Ni in the Li layers. The new ordered crystalline structure presents properties in nice agreement with the experimental data and when delithiated, can explain the driving force for the observed H1 to M phase transition.

As it is shown here the issue of the choice of the unit cells for calculating the energy of different crystal structures of the LNO layered system is not peregrine since their formation energies and their relative values are usually used to calculate the phase diagrams and the voltage-capacity curves through the cluster-expansion (CE) method [6, 22]. In this method the energies of a large number of Li$_x$NiO$_2$ ordered crystal structures in small unit cells are calculated and fit to extract cluster (pairs, triplets, quadruplets formed by lattice points) contributions to the energy. To each cluster, in a given ordered crystal structure, is associated a "correlation function" calculated associating to each lattice point a "spin" value dependent on the actual ion occupying that lattice point. This paper shows that, at least in the case of LiNiO$_2$, the choice of the basic cells needed to calculate the input total energies for CE may have a large influence, not only on the determination of ground structures and phase diagrams, but also on the structural, electronic, and magnetic properties associated to the input structures.

Moreover, since LNO suffers from the above mentioned structural instabilities and maintaining the structural stability during cycling is critically important to achieve a long-term cycling performance, the new crystalline structure is used to study the properties of Mn modified LNO. Indeed, the instability of LNO has originated research efforts aiming at stabilizing the material through the addition of other metal cations. Other layered oxides with general LMO formula have been synthesized in which M stays for other transition metals inserted in partial substitution of Ni in order to reduce the quantity of Cobalt. [23, 24]. Among these promising layered oxide compounds is nickel-rich NMC where N stays for Nickel, M for Manganese, and C for Cobalt. To eliminate Cobalt, layered Li$_x$Ni$_{(1-y)}$Mn$_y$O$_2$ (LNMO) solutions have been synthesized and tested. A few papers have focused on the LNMO material system. Layered Manganese-substituted LiNi$_{(1-x)}$Mn$_x$O$_2$ solid solutions can only be formed when $x$ is lower than 0.5 and have the same α-NaFeO$_2$ layered structure of $R\bar{3}m$ LNO in the nickel-rich regime ($x \leq 0.5$). For $x \geq 0.5$ a phase transition to a cubic spinel symmetry occurs. In most LNMO samples some Li is found in the transition metal layer, Li$_{TM}$,, so the solution is indicated as Li[Ni$_x$Li$_{(1-2x)/3}$Mn$_{(2-x)/3}$]O$_2$ ($0 \leq x \leq 0.5$). Manganese incorporated in pristine LNO, Li(Ni$_{(1-x)}$Mn$_x$)O$_2$ has been found in the tetravalent Mn$^{4+}$ oxidation state [25, 26], so that more Ni$^{2+}$ cations are supposed to be generated to conserve the charge, representing the fully lithiated solid solutions as Li(Ni$^{2+}_x$Ni$^{3+}_{(1-2x)}$Mn$^{4+}_x$)O$_2$. [27]. Obviously, this assumption stems from the assumption that in lithiated LNO all the Ni cations are Ni$^{3+}$, assumption which is challenged in the present paper.

The presence of Mn was also found to induce an increase of the Li/Ni cation mixing. The lithium ions in the transition metal layer, Li$_{TM}$, are electrochemically active resulting in the creation of tetrahedral sites for lithium occupancy when, during cathode charging, Li$_{TM}$ are removed [28]. First principles calculations have shown that the Li/Ni exchange promoted by Mn tends to increase the activation energy for Li ion diffusion [29]. Other authors [30] pointed out that the formation energy of an extra Ni defect in the Li layer in LiNi$_{0.5}$Mn$_{0.5}$O$_2$, increases with the presence of the high valence Mn$^{4+}$ dopant. This fact, together with the observation that the presence of Mn$^{4+}$ reduces Ni$^{3+}$ into Ni$^{2+}$, thus reducing the lattice distortion due to the Ni$^{3+}$ JT effect, is thought to lead to the observed stabilization of the crystal structure.

In this paper we model pristine layered LNMO (e.g. without Li$_{TM}$ or other kind of defects) using the large optimized supercells of pristine LNO of $R\bar{3}m$ space group. The aim is to study the evolution of the structural and electronic properties of the solid solutions versus Mn content and to investigate the effects of Mn introduction, in substitution of Ni, on the structural and electronic properties of the delithiated structures. Our study confirms that Mn ions are predominantly in the $Mn^{4+}$ state and, thus, do not participate in the redox reactions occurring during delithiation. They have, however, a role in determining the oxidation states of the Ni ions, and, thus, of the $Ni^{2+}/Ni^{3+}$ and $Ni^{3+}/Ni^{4+}$ redox couples which determine the battery charge and discharge potentials, and the intercalation voltage curves.

## 2. COMPUTATIONAL METHOD

Total energy calculations have been performed within the spin-polarized Density Functional Theory (DFT) approach using the plane-wave pseudopotential method as implemented in the Quantum Espresso Package [31]. The exchange and correlation density functional is expressed through the GGA approximation of Perdew-Becke-Ernzerhof (PBE) [32]. We use norm conserving (NC) pseudopotentials and all plane waves with energy less than 80 Ry are included in the basis set. The very high cutoff of 80 Ry assures a good convergence of the total energies. The Hellman-Feynman forces were converged to 0.005 eV/Å. The rotationally invariant scheme proposed by Dudarev et al. [33] was used to include an energy term based on the Hubbard U model, facilitating electron localization in the transition metal d orbitals. We chose U = 3 eV since using this value, the calculations predicted structural, magnetic and electronic properties of $LiNiO_2$ and $LiMnO_2$ in the rhombohedral ($R\bar{3}m$), monoclinic ($C2/m$), spinel ($Fd\bar{3}m$), and orthorhombic ($Pmmn$) crystal structures in reasonable agreement with the available experiments. We checked also other values of U and found that the physical interpretation of the results was not affected substantially. As shown in Fig. 1, the only relevant effect of the value of U on the Total Density of States (TDoS) of rhombohedral layered $LiNiO_2$ in its primitive unit cell, is to push upward the energy of the empty states and, in the specific case, to widen the gap between the spin down bands. However, in this paper we are interested in ground state properties while experimental energy gaps are better predicted by excited state theories. Evidence that the results of the calculations do not depend strongly on the U value was also given in literature [34].

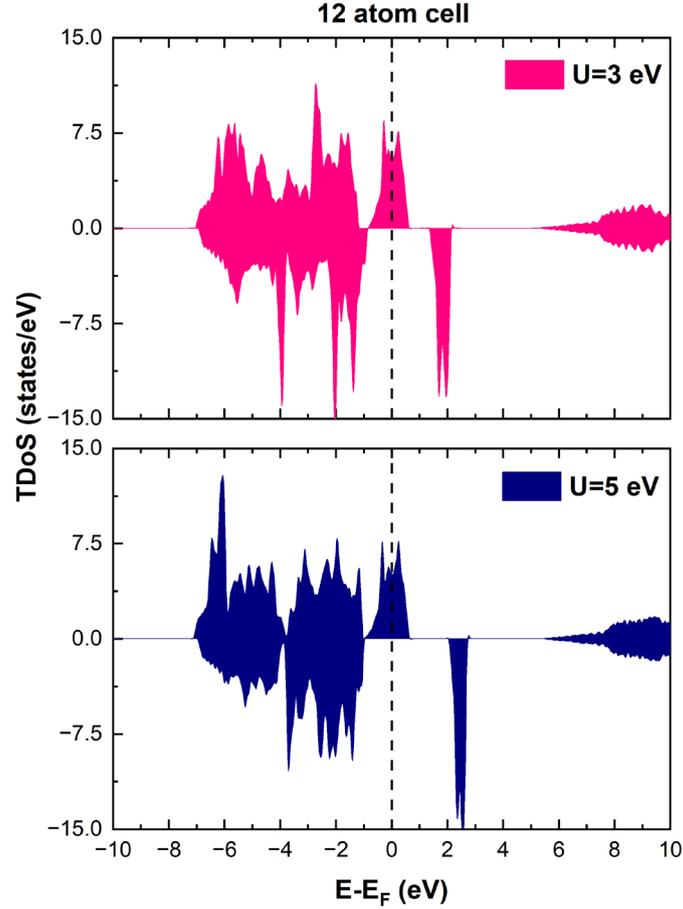

Fig. 1. Effect of U on the TDoS of LNO.

The calculation of the mixing energy against phase separation in a completely lithiated and a completely delithiated phases, was used to compare the stability of structures on the same lattice having a different number of Li vacancies. The formula used to calculate it is:

$$E_{Mixing} = E(Li_xNiO_2) - xE(LiNiO_2) - (1-x)E(NiO_2) \qquad (1)$$

where, E (LiNiO$_2$) is the total energy of the fully lithiated structure, E (Li$_x$NiO$_2$) is the total energy of the structure with Lithium concentration $x$; E(NiO$_2$) is the total energy of the fully delithiated structure. We consider only the internal energy since, as usually assumed in literature, the enthalpic and entropic contributions can be neglected when considering the systems at low temperatures.

Analogously the stability of Li$_x$Ni$_{1-y}$Mn$_y$O$_2$ is assessed by defining the mixing energy as:

$$E_{Mixing} = E(Li_xNi_{1-y}Mn_yO_2) - x(1-y)E(LiNiO_2) - xyE(LiMnO_2) - (1-x)(1-y)E(NiO_2) - (1-x)(y)E(MnO_2) \qquad (2)$$

Here, E(Li$_x$Ni$_{1-y}$Mn$_y$O$_2$) is the internal energy of the Mn/Ni substituted structure with a concentration $x$ of Lithium, [(1-x) Li vacancies] in the Li layers, and a concentration $y$ of Mn [(1-y) of Ni] in the transition metal layers. E(LiNiO$_2$), E(LiMnO$_2$), E(NiO$_2$) and E(MnO$_2$) are the energies of the fully lithiated LNO and LMO, and the fully delithiated NO and MO, respectively. All the structures are considered in the same crystal structure. This choice for the stability assessment of $Li_xNi_{(1-y)}Mn_yO_2$ was previously used [35] to investigate how the stability of different arrangements of the two kinds of transition metal ions depends on their mutual interaction, rather than to determine the absolute stability of the compounds from constituent elements.

Ferromagnetic alignment of Ni spins is adopted throughout all the calculations since it was shown that the spin interaction energies are smaller than those between different structural phases and a ferromagnetic ordering corresponds to the ground state of LNO and Li-rich and Mn-poor LMNO [21, 36].

## 3. Fully Lithiated LiNiO2

### 3.a Structural Properties

LiNiO$_2$ can exist as a layered oxide with an arrangement of oxygen atoms forming a cubic closed packed structure. In the crystalline rhombohedral $R\bar{3}m$ space group the transition metal and lithium ions occupy octahedral sites. In the O3 stacking, shown in Fig. 2, the primitive 12 atom cell, Fig. 2 (a), contains three formula units, consisting of parallel planes of lithium, oxygen, nickel, oxygen atoms. The Li ions diffuse, intercalate and de-intercalate between the planes of NiO$_6$ octahedra. Typically, the rhombohedral space group is represented in the corresponding hexagonal (H) cell with lattice parameters $a$, $b$, and $c$ (orthogonal to $a$ and $b$).

We performed calculations using simulation cells obtained increasing the $a$ and $b$ in-plane lattice parameters of the hexagonal cell twice and four times, containing 48 and 192 atoms, respectively. The unit cells are shown in Fig. 2. The structural relaxation of the larger cells has been performed starting from the cell parameters and atom positions of the optimized primitive 12 atom cell.

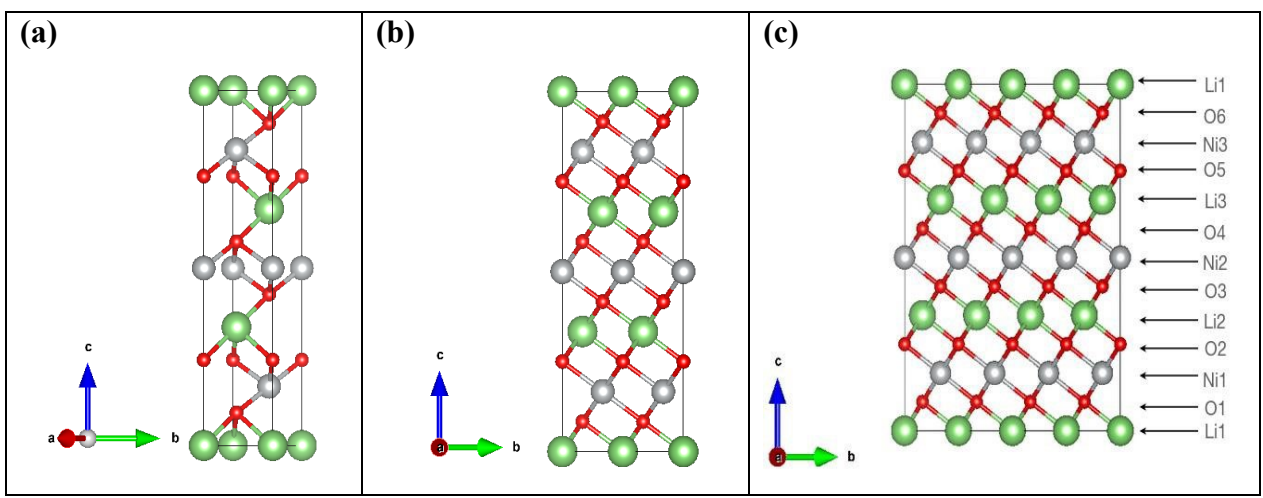

Fig. 2. (a), (b), (c) Ball and stick representation of LNO in the $R\bar{3}m$ space group with 12, 48 and 192 atoms unit cells, respectively.

In the 12 atom (1×1×1) and 48 atom (2×2×1) cells the NiO$_6$ octahedra do not change appreciably but in the 192-atom unit cell (4×4×1) the structural relaxation leads to octahedra size disproportionation with expanded and compressed octahedra having different Ni-O bond lengths. The octahedra show also a small JT distortion. Also, the cell lattice parameters change with the cell dimension as shown in Table 1. The expanded and compressed octahedra are related to a disproportionation of the Ni ions oxidation states as shown in Table 2, where the bond lengths of the different kinds of octahedra in the 192-atom supercell are also reported.

Table 1. Optimized values of structural parameters in fully lithiated LiNiO$_2$.

| No. of atom cell | | Lattice Parameters (Å) | | | Other theory | Expt. |
|---|---|---|---|---|---|---|
| | | a | b | c | | |
| $R\bar{3}m$ | 12 | 2.89 | 2.89 | 14.25 | (2.84, 14.29) [37], (2.89,14.20) [37], (2.90, 14.18) [38], (2.88,14.27) [39] | (2.88, 14.18) [15], (2.87,14.21) [14], (2.88, 14.19) [40], (2.90, 14.2) [41] |
| | 48 | 2.86 | 2.86 | 14.05 | | |
| | 192 | 2.87 | 2.87 | 14.01 | | |
| C2/m | 8 | 4.96 | 2.96 | 5.74 | (5.16, 2.79, 5.13) [7] | |
| | 192 | 9.83 | 11.71 | 5.74 | | |

Table 2. Calculated Ni–oxygen bond lengths in the different atom cells.

| System | No. of atoms in cell | Metal-Oxygen Bond | Bond length (Å) |
|---|---|---|---|

| | | | | | |
|---|---|---|---|---|---|
| $R\bar{3}m$ | 12 | $Ni^{3+}$-O | 1.98×6<br>Δ=0.0 | 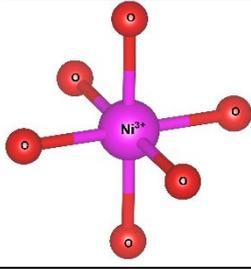 | |
| | 192 | $Ni^{2+}$-O | 2.075×2 (A)<br>2.019×2 (C)<br>2.009×2 (D)<br>2.034 (Average)<br>Δ=0.06 | 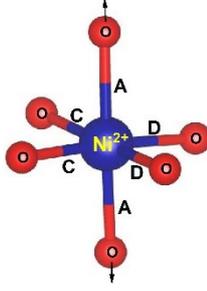 | 2.026×4 (A)<br>2.040×2 (C)<br>2.033 (Average)<br>Δ=0.014 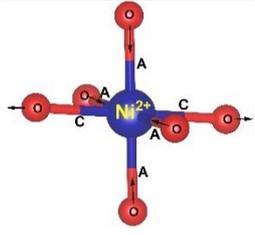 |
| | | $Ni^{4+}$-O | 1.924 (A)<br>1.88 (B)<br>1.902 (C)<br>1.89 (D)<br>1.899 (Average)<br>Δ=0.044 | 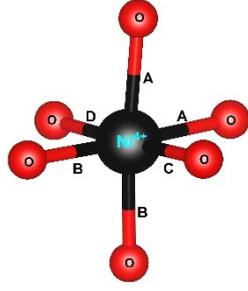 | |
| C2/m | 8 | $Ni^{3+}$-O | 2.14 (A)<br>1.904 (B)<br>1.921 (C)<br>1.988 (Average)<br>Δ=0.236 | 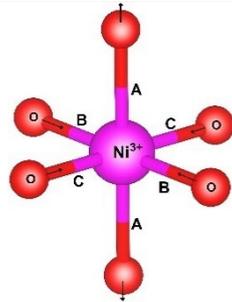 | |

*Here Average is the arithmetic average, and Δ is difference between highest and lowest values.

A phase consisting in a mixture of expanded, compressed, and JT distorted octahedra corresponding to $Ni^{2+}$, $Ni^{4+}$ and $Ni^{3+}$, respectively, was assumed from experimental neutron pair distributions which could agree with most of the existing experimental data. The authors claimed that such a disordered metastable "charge-glass" phase could be stabilized by entropic effects [21].

Here, we have found a structure consisting of expanded and compressed octahedra with Ni charge disproportionation by optimizing the $R\bar{3}m$ O3 structure in the 4×4×1 supercell, while the monoclinic structures, of which the C2/m is one representative, have JT distorted octahedra with four short and two long Ni-O bonds and $Ni^{3+}$ cations at their center (see Table 2). Indeed, there are two modes of

JT distortions occurring in layered LNO, Q2 and Q3, discussed in detail in literature [42]. In the *C2/m* structure, only the Q3 mode takes place, leading to a collinear JT distortion, where all the long bonds are similarly oriented. We define a JT distortion parameter Δ which is the average difference between the longest Ni-O bonds and the shortest Ni-O ones. The values obtained for the hexagonal and monoclinic structures are given in Table 2.

This result is the first one reporting charge disproportionation and relative $NiO_6$ octahedra size disproportionation in the fully lithiated O3 layered hexagonal crystal structure.

Our results are in qualitative agreement with the neutron diffraction data [16] where the peak broadening was found more consistent with continuous distributions of Ni-O bond lengths due to nonuniform distortions than with the two different fixed Ni-O bond lengths expected in the case of a monoclinic lattice. Our calculated average long and short Ni-O bond lengths are 2.03 and 1.90 Å which nicely compare with the experimental 2.05 and 1.93 Å [16] and with the X-ray absorption data finding Ni-O bonds 2.08 and 1.91 Å long, respectively [43].

### 3.b Electronic and Magnetic Properties

The electronic structure is discussed by inspecting the TDoS and partial density of states (PDoS) of the considered structures. In Fig. 3 (a), we see that, using the 12 atom cell for the $R\bar{3}m$ crystal structure, the spin up TDoS has an half occupied band at the Fermi level, while the spin down TDoS has an electronic gap. In this case the Ni atoms are at the center of similarly perfect $NiO_6$ octahedra (see Table 2). Since the crystal field splits the five *d* states into a three-fold degenerate $t_{2g}$ state and a two-fold degenerate $e_g$ state, the Ni states should be all in a $Ni^{3+}$ oxidation state which adopts the low spin configuration $t_{2g}^6 (|\uparrow\downarrow|\uparrow\downarrow|\uparrow\downarrow|)\ e_g^1 (|\uparrow\ |\ |)$, where the $t_{2g}$ states are occupied leaving one electron in the doubly degenerate $e_g$ state at higher energy. In the crystal the Ni *d* states are hybridized with the Oxygen *p* states, thus the half-occupied feature at the Fermi level corresponds to the antibonding $e_g^*$ band. The unpaired $e_g$ Ni electrons occupy half of the $e_g^*$ spin up band while the $e_g^*$ spin down band remains empty. This behavior is reflected in the A peak which corresponds to the $e_g^*$ spin up orbitals and is half occupied as shown in Fig. 3 (b). The B peak corresponds to the spin down $e_g^*$ states.

The $Ni^{3+}$ cations have an estimated magnetic moment of 1 $\mu_B$ (based on the calculated difference of a Löwdin partition of the up and down spin charge on atoms) and the preferred phase is ferromagnetic. The estimated local magnetic moments on the transition metal cations are reported in Table S1 in the Supporting Information (SI).

These results compare well with calculations reported in the literature, which, invariably use the 12-atom cell to describe the fully lithiated $R\bar{3}m$ LNO [37, 39, 44].

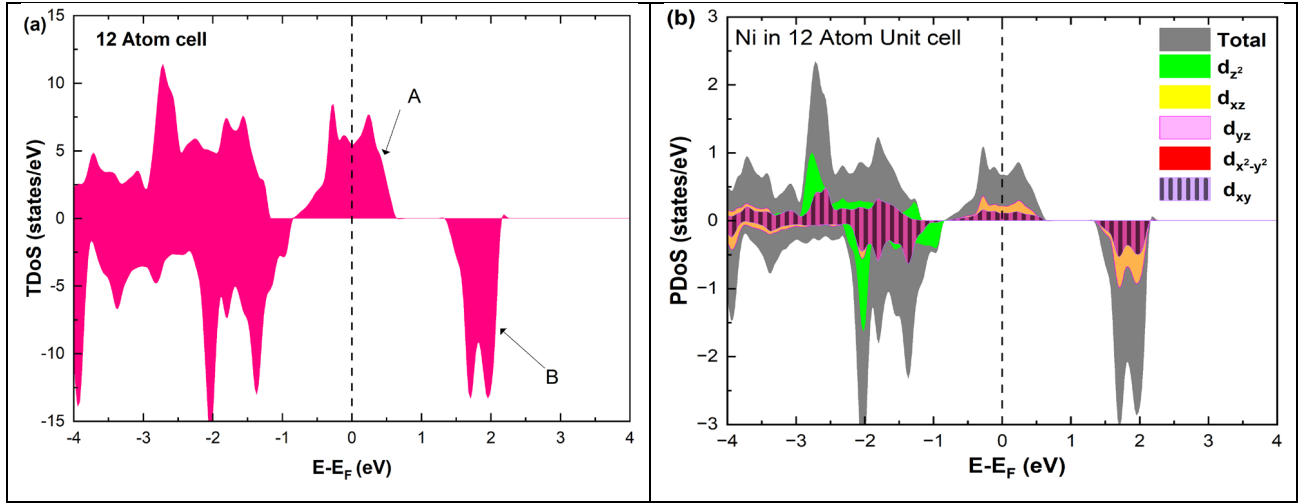

Fig. 3 (a) Total density of states, and (b) Atom centered Partial density of states in 12 atom fully lithiated LiNiO$_2$.

The $Ni^{3+}$ cation having only one electron in the $e_g^*$ orbitals is J-T unstable. However, the distortions cannot take place in the small 12 and 48 atom cells, thus the octahedra have equal Ni-O bond lengths (see Table 2).

As seen above a structural distortion takes place instead in the 192-atom supercell, and, as a consequence, the A peak feature at the Fermi Level in the TDoS is significantly changed, as shown in Fig. 4 (a). The spin up A band splits into a mostly occupied $A_1$ peak and an empty $A_2$ peak, hinting at the formation of a band gap, and similarly the spin down B peak splits into $B_1$ and $B_2$. The states just below the Fermi energy show a predominant oxygen and, at a lesser extent, nickel character as shown in Fig S1 (SI). These hybridized (O $p$-Ni $d$) energy states are of particular interest, as these are the states from which electrons are removed when the delithiation process occurs.

Analyzing the partial density of states, we find that there are two kinds of Ni ions as shown in Fig. 4 (c, d). The $e_g^*$ spin up state of the first one is almost fully occupied while that of the second one is mainly unoccupied. This situation is reflected in their different magnetic moments reported in Table S1 (SI). Ni ions have two oxidation states: $Ni^{4+}$ with the $e_g^*$ spin up level completely empty and magnetic moment zero, and Ni ions close to the $Ni^{2+}$ oxidation state with an almost full $e_g^*$ spin up band and an estimated magnetic moment, about 1.4 $\mu_B$, much larger than 1 $\mu_B$, the one associated to $Ni^{3+}$. We name these second Ni ions simply as $Ni^{2+}$. The features of the PDoS can be assigned neatly to these two different Ni cations: the $A_1$ peak is the occupied part of the $3d$-$e_g$ orbital of $Ni^{2+}$ (Fig.

4(c)), as well as the empty peak B$_2$, while the empty peaks A$_2$ and B$_1$ belong to Ni$^{4+}$ (see Fig. 4(b)). The existence of Ni$^{4+}$ ions in LNO has been calculated previously only after some Li vacancies are present [45] and never for the fully lithiated ideal structure. This was due to the use of the smaller unit cells to describe LNO in the layered $R\bar{3}m$ space group. Such charge disproportionation into Ni$^{2+}$ and Ni$^{4+}$ was calculated before only for the monoclinic P2/c structure [34] and in monoclinic C2/m LNO only after one additional Ni ion was inserted in the Li plane [30].

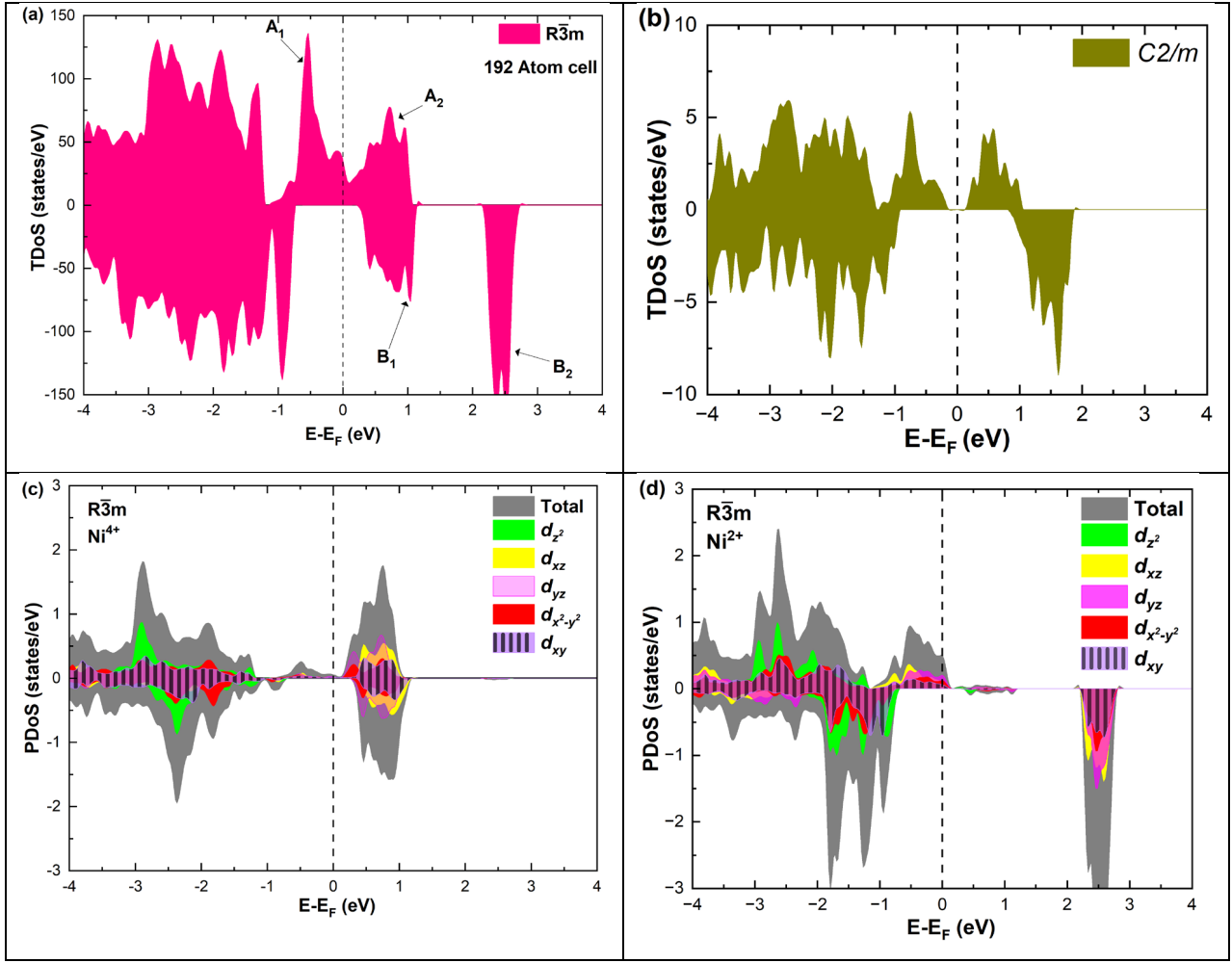

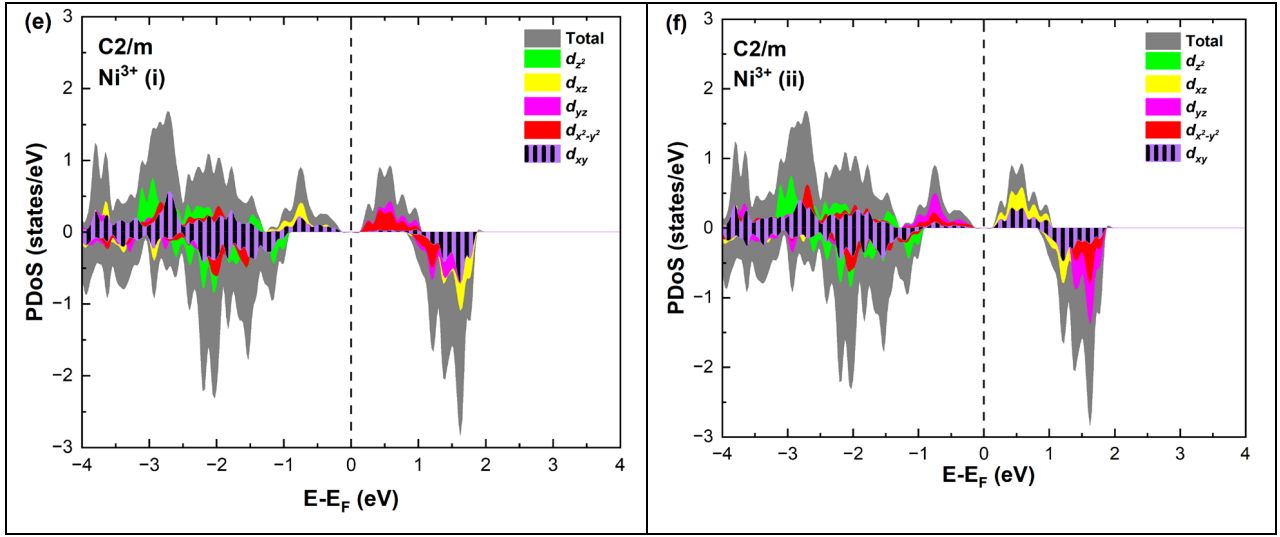

Fig. 4 TDoS in (a) $R\bar{3}m$ and (b) $C2/m$; Nickel atom centered partial density of states in $R\bar{3}m$ (c, d) and in $C2/m$ (e, f) in fully lithiated $LiNiO_2$.

The distorted $NiO_6$ octahedra of the 192-atom supercell lead to the breaking of the $e_g$ and $t_{2g}$ orbital degeneracies. The $Ni^{4+}$ cations, Fig. 4 (c), where both the spin up and spin-down $e_g^*$ bands are empty, are at the center of the compressed octahedra with short Ni-O bond lengths (see Table 2). In this case the (Ni $d$ –O $p$) hybridization increases and the antibonding $e_g^*$ bands are shifted at higher energies remaining empty. The $Ni^{2+}$ cations, instead, are at the center of the expanded octahedra having longer Ni-O bond lengths, Fig. 4 (d). The hybridization in this case decreases and the $e_g^*$ spin up bands shift down in energy becoming mostly occupied. The presence of $Ni^{2+}$ cations in an ideally layered LNO would explain why it is so difficult to obtain samples without Ni in the Li layer given the very low energy barrier for $Ni^{2+}$ to replace $Li^+$.

The TDoS of the $C2/m$ structure showed in Fig. 4 (b), depicts band gaps in both spin channels as found by other authors [7]. The PDoS of the two Ni atoms in the primitive cell, shown in Fig. 4 (e, f), have the same 0.25 eV band gap which is due to the splitting of the $e_g^*$ bands for the JT effect. There is no charge disproportionation in this case, so the Ni oxidation states are $Ni^{3+}$. For both $Ni^{3+}$ ions only the spin up electron goes in the lowest energy JT split $e_g^*$ band below the Fermi energy, while the other spin up band and both the spin down bands, closer in energy, are empty.

In order to give a clear understanding of the difference in the electronic configurations associated to the $e_g^*$ spin up bands in the hexagonal 192 atoms supercell and in $C2/m$, starting from the undistorted octahedra case of the $R\bar{3}m$ structure in the primitive 12 atoms cell, we show in Fig. 5 a scheme of the different splittings and occupations of the spin up $e_g$ states on the Ni atoms. The $e_g$ spin down orbitals are higher in energy and unoccupied.

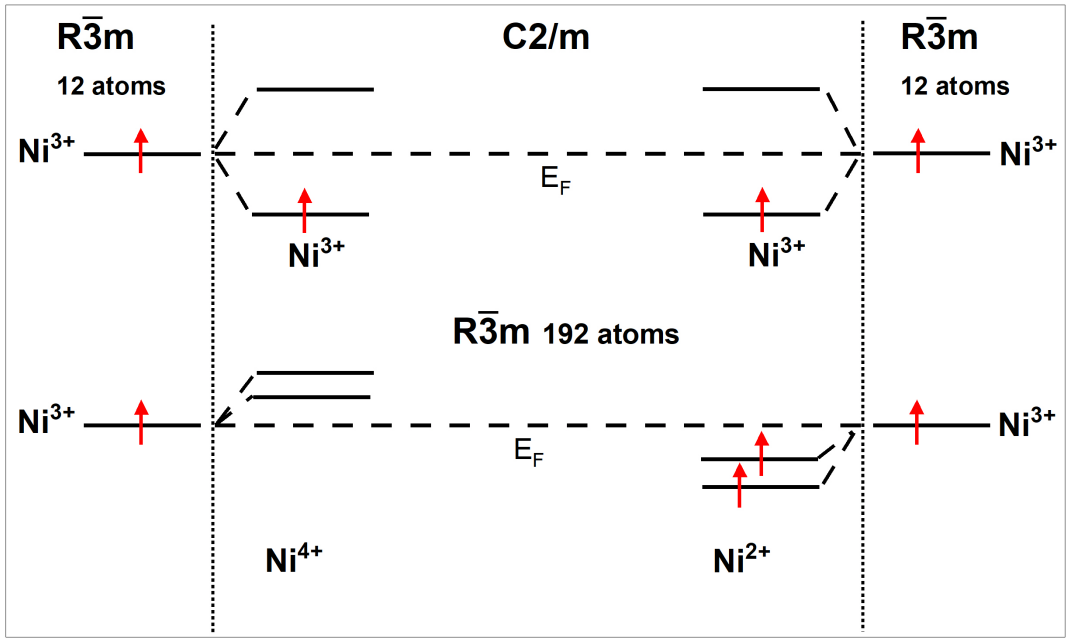

Fig. 5. Schematic diagram of Ni $d$ $e_g$ spin up states in the hexagonal 192 atoms structure and in the $C2/m$ structure.

The arrangement of the $Ni^{4+}$ and $Ni^{2+}$ ions form in-plane ordered patterns with 9 $Ni^{2+}$ and 7 $Ni^{4+}$ as shown in Fig. 6. If the Ni ions were all in the $Ni^{3+}$ oxidation states then there will be 16 electrons in the $e_g^*$ bands per plane, one per Ni atom. If $Ni^{4+}$ ions have no electrons in $e_g^*$ then the 9 $Ni^{2+}$ need to have a total of 1.78 electrons in $e_g^*$. Assuming that no charge is transferred to other bands, their average oxidation state is actually +2.22. The $Ni^{2+}$ ions form two kinds of hexagons, one regular, the other irregular with 9 ions on the perimeter. The hexagons surround three $Ni^{4+}$ ions. The symmetry of the structure belongs to the hexagonal space group, with 12 symmetry operations, as the structure described with the 12 atom unit cell, but the structural, electronic and magnetic properties are quite different.

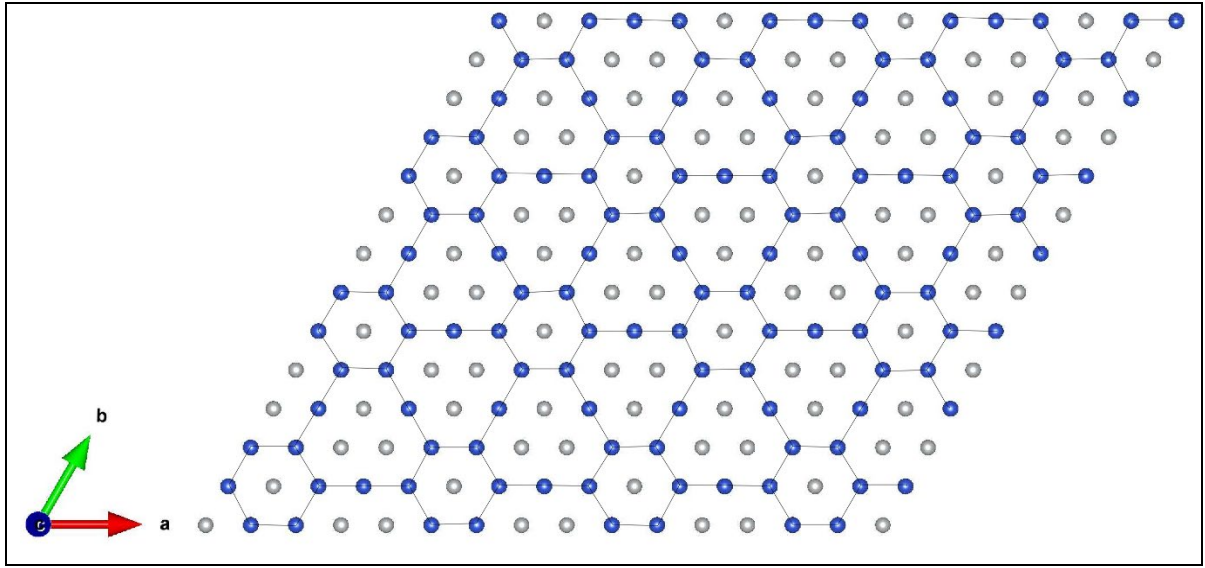

Fig. 6. Pattern formed by the $Ni^{4+}$ (Grey) and $Ni^{2+}$ (Blue) cations in the transition metal layers of the 192 atom supercell.

Our calculations have shown the strong dependence of the physical properties of the hexagonal fully lithiated LNO on the choice of the unit cell. Since the octahedra have common edges, they cannot distort independently, thus, it is quite possible that the use of even larger unit cells could allow more freedom for octahedral distortion which could lower the energy even further and lead, possibly, to the opening of a gap between the spin up states as in the case of the *C2/m* monoclinic point group, in accordance with the experimental observation of a 0.5 eV gap for the almost stoichiometric LNO in the hexagonal phase [46]. However, gaps are excited state properties and not always can be correctly predicted by DFT methods. The total energy per atom of LNO in hexagonal 192 atom cell, and *C2/m*, in its primitive 8 atom unit cell are almost the same, within the accuracy error of the calculations.

## 4. Partially Lithiated LNO

**4.a Li Vacancy Mixing Energies and Structural Properties**

The structural models of the delithiated $Li_xNiO_2$ structures are obtained starting from the relaxed hexagonal supercell with 192 atoms used as a template. We know from the literature that during the charge and discharge cycles the delithiated systems undergo various phase transitions (H1, M, H2, H3) at certain values of Li concentration *x*, maintaining however the layered structure. [3, 4, 12, 13, 14]. Li cations are removed from only one of the three Li planes characterizing the O3 stacking, (single layer structures), or from all the three planes (multilayer structures). For the single layer structures eight different Li vacancy configurations were calculated realizing four concentrations $x$ = 0.75, 0.625, 0.5, 0.25. These configurations named 1c and 1b ($x$ = 0.75), 2a and 2b ($x$ = 0.625), 3a and 3b ($x$ = 0.5), and 4a and 4b ($x$ = 0.25) are shown in Fig. 7 (a).

In the case of the multilayer structures there are numerous ways in which the Li ions can be removed from each layer and many ways in which the layers can be stacked for each Li concentration. Here, we have considered only some of the possible stacking combinations since our aim is to obtain a general overview of the trends in the properties of the delithiated structures as a function of the Li content $x$. Some of the structures are shown in Fig. 7 (b). We have generated different vacancy configurations for each Li layer (a sequence), then we have stacked them in different ways, sometime just repeating the same sequence in all the three Li planes, other times using different sequences in the three planes. We have generated 22 different configurations covering Li concentrations $x$ = 0.916, 0.833, 0.75, 0.667, 0.58, 0.5, 0.41, 0.33, 0.25, 0.167, 0.083. For each Li content we compared the energies of two configurations. The two configurations were chosen so as to have the Li vacancies close to one another or far apart. A complete list of the structures is reported in Table S2. Li vacancy configurations have been used by other authors [5, 47, 48] using smaller monoclinic unit cells, in which only two oxidized Ni states ($Ni^{3+}$ and $Ni^{4+}$) were found as a consequence of delithiation.

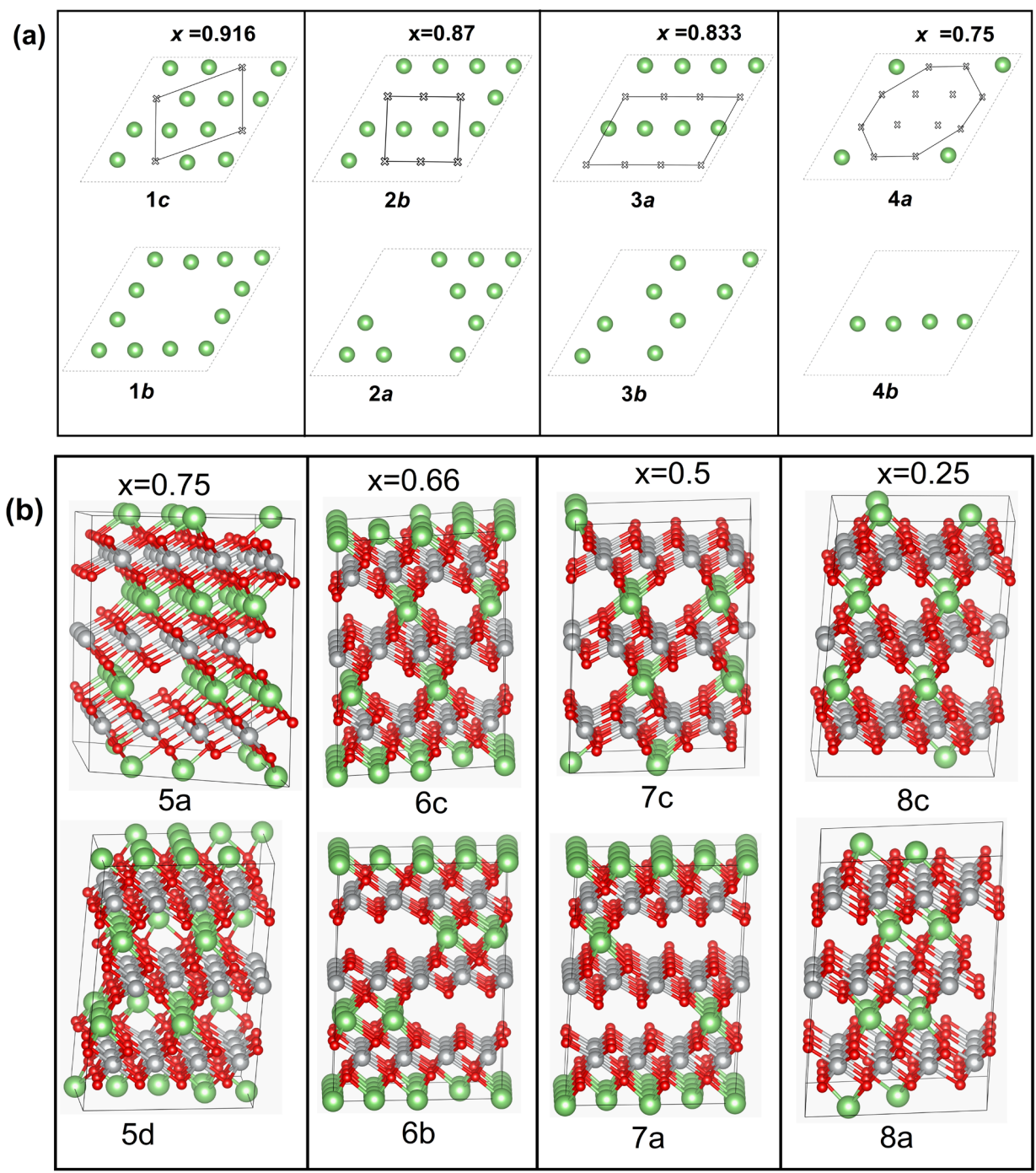

Fig. 7. Delithiated structures in the 192 atom cell: (a) Single layer structures: 1c and 1b (8.33% Li vacancies in total), 2b and 2a (12.5%), 3a and 3b (16.67%), 4a and 4b (25%); (b) Multilayer structures: 5a and 5d (25% Li vacancy), 6c and 6b (33% Li vacancy), 7c and 7a (50% Li vacancy) and 8c and 8a (75% Li vacancy).

The mixing energies of these partially lithiated $Li_xNiO_2$ are calculated using Eq. (1) and only those whose energy was found lowest are plotted in Fig. 8 (in the following we will name these structures as "most stable" but, since we have not considered all the possible vacancy configurations over the

48 Li sites in the 192-atom supercell, the structures are just the most stable among those calculated). Nevertheless, even within this limitation, interesting trends are found. For the same Li content, the multilayer structures have in general a lower energy than the single layer ones, showing that delithiation occurs preferably removing Li simultaneously from all the layers.

At high values of $x$, the more stable configurations are characterized by an ordering of the Li vacancies which tends to maximize their mutual distance. This result suggests the existence of a repulsive interaction between the Li-vacancies which is decaying with the distance. At lower $x$, when the vacancies need to be closer, the most stable configurations present rows of vacant and occupied Li sites. These chain-like arrangements of Li atoms and Li vacancies seem to characterize favorable Li vacancy arrangements. This can be due to the strong interaction between Li cations of different planes along the direction of the Li-O-Ni-O-Li chains [5] that in presence of Li vacancies will induce the JT distortion of the $NiO_6$ octahedra. Indeed, the presence of Li ions and vacancies in ordered interlayer chains favor the collinear formation of the JT distortions with a smaller lattice strain since channels are formed, as shown in Fig. 7.

The optimized lattice parameters of the delithiated structures are reported in Table S3.

In Fig. 8 we show, together with the mixing energies of the lowest energy delithiated configurations also the energies of the fully lithiated hexagonal and $C2/m$ LNO structures, both calculated using the primitive unit cells (12 atoms for $R\bar{3}m$ and 8 atoms for $C2/m$) and 192 atom supercells. The energies are all relative to the energy of the hexagonal 192 atom supercell. We can see that the energy of the hexagonal structure decreases with the atomic position optimization passing from the primitive cell to the supercell much more than the energy of the monoclinic structure, showing thus a more efficient structural relaxation. The relaxed structures in the 192 atom supercells are much closer in energy, 8.75 meV/atom.

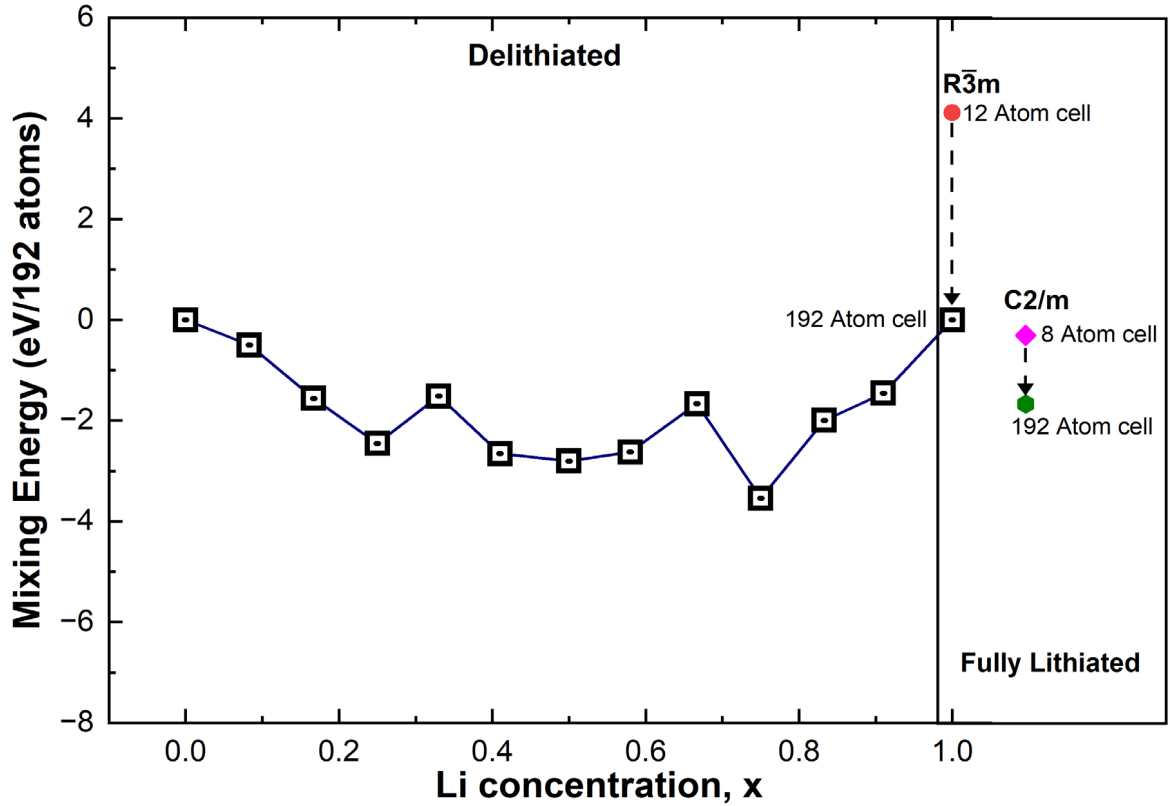

Fig 8. Mixing energies of $Li_xNiO_2$.

In Fig. 9 we report the variation of the lattice parameters $a = b$ and $c$ versus Li concentration $x$. The lattice parameters of the lowest energy structures are reported for each $x$. The decreasing trend in $a$ is due to the progressive increasing of the number of $Ni^{4+}$ ions, which have a smaller ionic radius (0.46 Å) than $Ni^{2+}$ (0.69 Å) and $Ni^{3+}$ (0.56 Å) and are associated with the compressed octahedra. This trend is consistent with the experimental results [16,43]. Sharp contraction in the $c$ vertical lattice parameter within the range of $1 < x < 0.80$ is due to the emergence of JT active $Ni^{3+}$ cations out of the $Ni^{2+}$ cations. The number of $Ni^{3+}$ cations and the correspondingly JT distortion parameter $\Delta$ increase, as shown in Table 3, and they are largest in the Li concentration region $0.50 < x < 0.75$ where the layered structure undergoes the observed structural phase transition from the hexagonal H1 to the monoclinic M phases and remains in the monoclinic structure (M) (see Fig. 9). The other dip in the $c$ axis for $x < 0.4$ is due to the dominance of $Ni^{4+}$ ions and to the low Li content, so the octahedra become compressed, the JT distortion reduces, and the interaction of the Ni ions with the O ions increases. The Li interlayer distance reduces as less Li is present and the Oxygen atoms do not repel anymore as the Oxygen electronic $-2e$ charge is now more localized inside the octahedra. The reduction of the JT distortion leads to a new phase transition back to an hexagonal phase. At high delithiation levels the H3 phase was found to be unstable, probably due to oxygen loss with electrochemically inactive NiO formation. This trend in the $c$ axis with Li content, similar to that calculated also for $Li_xCoO_2$, was obtained by other authors only after using the SCAN and SCAN+D3 functionals in place of the PBE+U functional used in this work. In that work PBE+U produced only an ever decreasing $c$ lattice parameter [49]. In our case the expected trend has been obtained using PBE+U with the larger supercell.

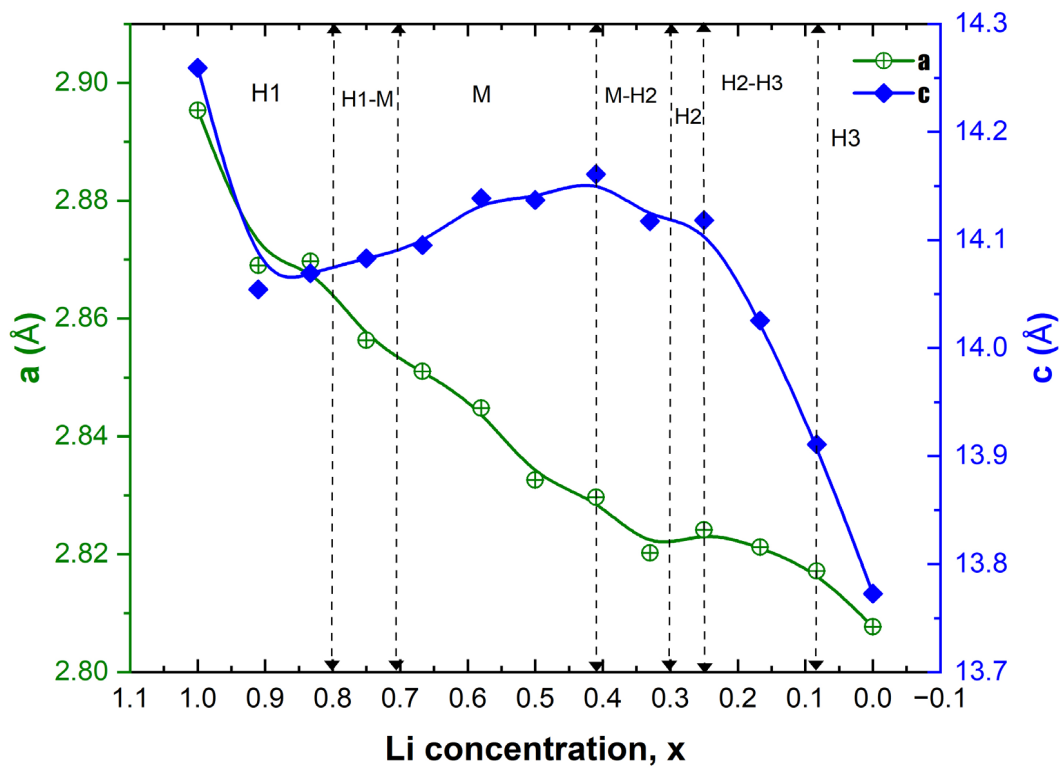

Fig. 9 Variation of the crystal lattice parameters, $a$ and $c$, as a function of Li concentration $x$. The vertical dashed lines correspond to the Li content at which phase coexistence and phase transition between the hexagonal H and the monoclinic M phases have been evidenced by XRD during the charge of the LNO cathode [14].

**4b. Electronic and Magnetic Properties**

Table 3 reports the number of $Ni^{2+}$, $Ni^{3+}$, and $Ni^{4+}$ ions present in the delithiated $Li_xNiO_2$ structures for $x$ = 0, 0.25, 0.5, 0.75, 1. The number of $Ni^{2+}$ ions decreases with $x$, while that of $Ni^{4+}$ increases. The number of $Ni^{3+}$ ions increases until $x$ = 0.5 then decreases. The active electrons are those in the $e_g^*$ orbitals and we can see that the number of these electrons are approximately 48 (54), 36, 24 (26), 12 (13), 0, (considering the nominal values of 2 and 1 for $Ni^{2+}$ and $Ni^{3+}$, respectively), where in the parentheses are the actual numbers compared with the nominal ones. These numbers indicate that roughly all the electrons lost by the cell in the delithiation are lost by the Ni cations.

Table 3. Number of different Ni cations and Δ parameter in Li$_x$NiO$_2$.

| Li concentration | No. of Ni cations | | | Jahn-Teller distortion parameter Δ in Li$_x$NiO$_2$ | | |
|---|---|---|---|---|---|---|
| $x$ | Ni$^{2+}$ | Ni$^{3+}$ | Ni$^{4+}$ | Ni$^{2+}$ | Ni$^{3+}$ | Ni$^{4+}$ |
| 1 | 27 | 0 | 21 | 0.06 | - | 0.04 |
| 0.75 (Chain) | 12 | 12 | 24 | 0.11 | 0.20 | 0.03 |
| 0.75 (Square) | 12 | 12 | 24 | 0.03 | 0.21 | 0.04 |
| 0.5 | 4 | 18 | 26 | 0.12 | 0.18 | 0.03 |
| 0.25 | 2 | 9 | 37 | 0.11 | 0.16 | 0.02 |
| 0 | 0 | 0 | 48 | 0 | 0 | 0 |

The different Ni oxidation states in the delithiated Li$_x$NiO$_2$ structures contribute to the electronic TDoS shown in Fig. 10, for $x$ = 0, 0.25, 0.50, 0.75, 1. The peaks and features within the energy range from -1 eV to 3 eV correspond, as for the fully lithiated case, to $e_g^*$ orbitals. To compare the evolution and shifts of the features we have chosen for all the partially delithiated structures similar configurations for the arrangement of the Li vacancies, in the specific the chain-like pattern, with one chain for $x$ = 0.75, two chains for $x$ = 0.50 (see for example Fig. 7 case 3a), and three chains for $x$ = 0.25.

In the case of $x$ = 0.75, the chain arrangement of Li vacancies is not the most stable. Thus, we compare in Fig. 11 the TDoS of the chain arrangement (Fig. 11 (a)) with the TDoS of the more stable square arrangement (Fig. 11 (b)). The two TDoS are qualitatively similar, since the spin up and spin down states fall in the same ranges of energy. The main difference stems from the different height of the peaks. Peaks are more pronounced for the chain arrangement. Thus, in the case of the chain arrangement the energies of the Ni-O $e_g^*$ orbitals fall preferably in narrower intervals while for the square arrangement their energies are more smoothly distributed over their energy range. Both delithiated configurations have the same number of Ni$^{2+}$, Ni$^{3+}$, and Ni$^{4+}$ cations (see Table 3). The chain arrangement has however three kinds of oxygen ions: 24 $O"$ ions, first neighbors to two Li vacancies, 24 $O'$ ions, first neighbors to one Li vacancies and 48 O ions bonded to three Li ions. These different kinds of Oxygen ions are also arranged in similar chains. The less charged $O"$ tend to bond strongly to Ni ions that become Ni$^{4+}$. This structure has more groups of similar Ni-O bonds than the "Square" pattern that has 24 O ions with no first neighbor vacancies and 72 $O'$ ions with only one first neighbor Li vacancy. The O ions are also arranged in a square pattern. In this case the different kinds and lengths of Ni-O bonds are more uniformly distributed over the structure. This situation is shown in Fig. 12, where in the case of the chain arrangement the groups of chain aligned $O''$-$Ni^{4+}$ bonds are clearly visible.

The main features in the TDoS of Fig. 10 are named $A_1$, $A_1'$, $A_2$, $B_1$, $B_1'$, $B_1''$, $B_2$. The A peaks are constituted of spin-up orbitals and the B peaks of spin-down orbitals. $A_1$, $A_1'$ are occupied, $A_2$ is empty. We can see that the occupied part of $e_g^*$ progressively reduces with the increase of the number of Li vacancies due to the reduction of available electrons and the $A_1$ peak height reduces until it disappears at $x = 0.25$.

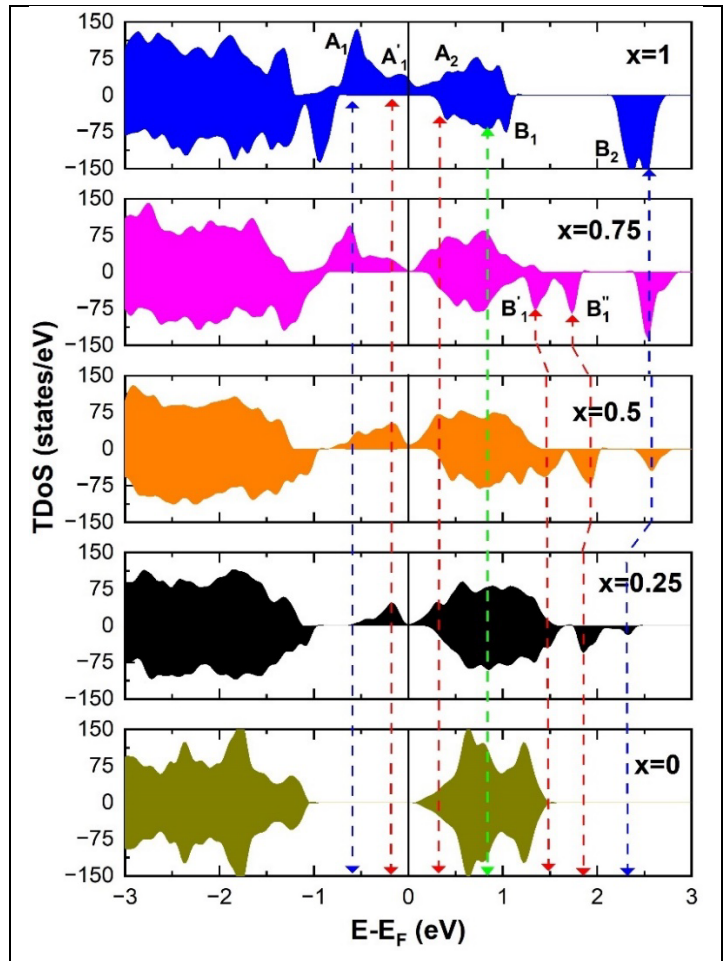

Fig. 10. Total Density of states (TDoS) of $Li_xNiO_2$

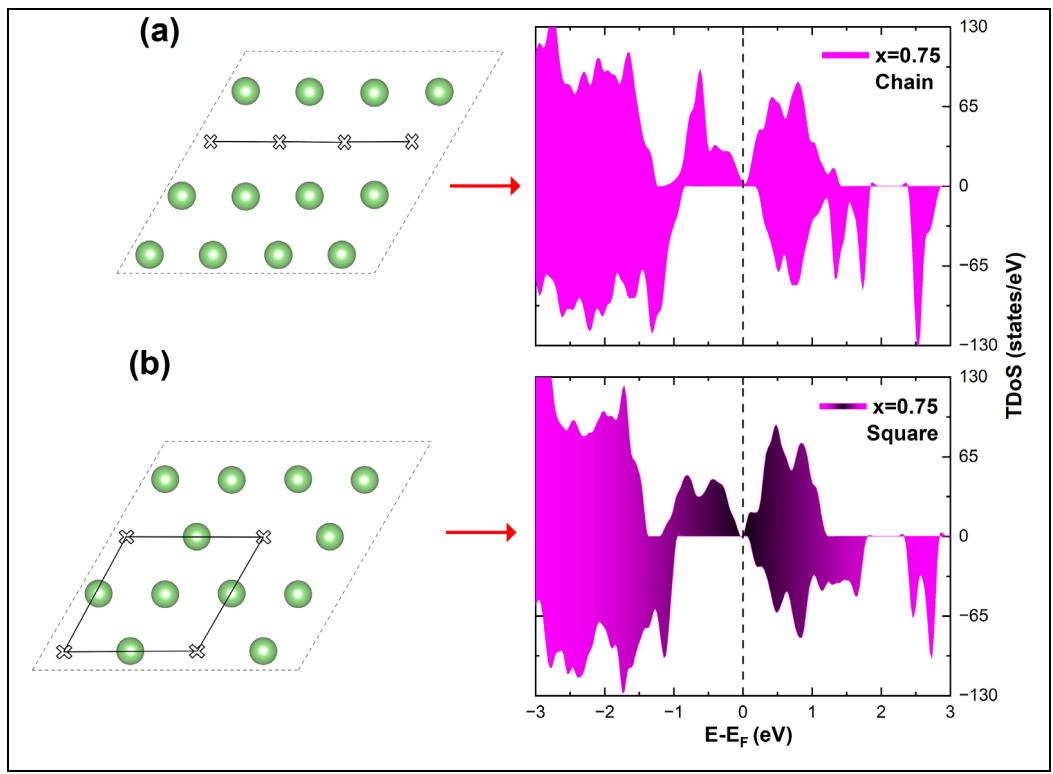

Fig. 11. Comparison of TDoS of two delithiated patterns at $x$ =0.75.

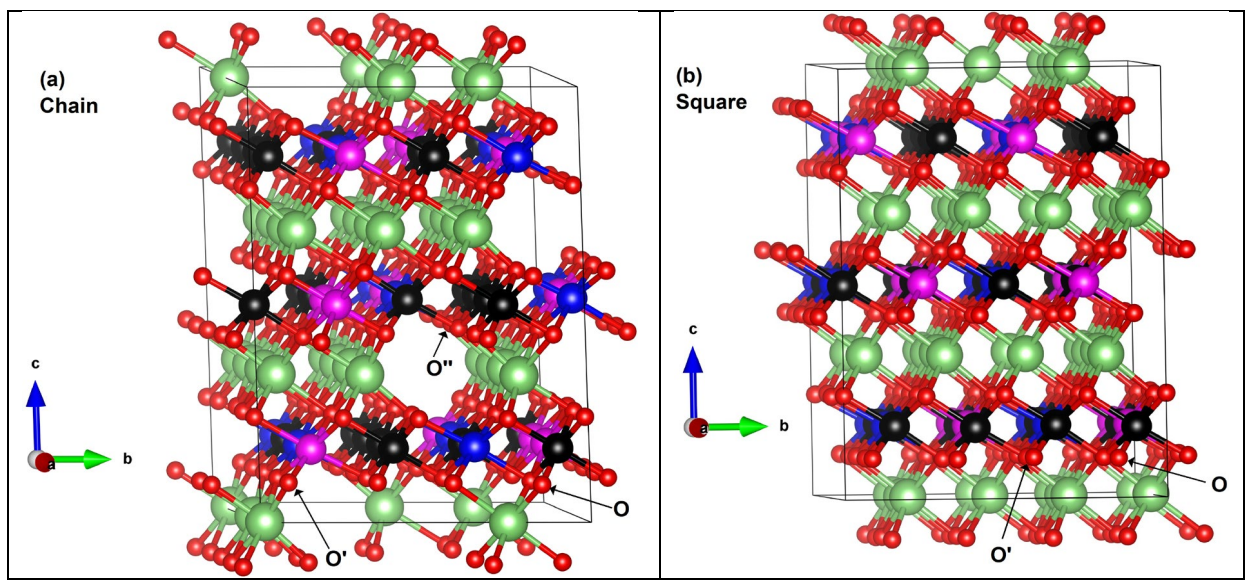

Fig. 12. Ball and stick representation of the two delithiated structures at $x$ = 0.75 showing the change of oxidation states of Ni in relation to the Li vacancies: (a) chain model, (b) square model. Li is represented by green balls, while Ni is represented by three different colors depending on the oxidation state. Black color defines $Ni^{4+}$, blue color $Ni^{2+}$ and magenta color $Ni^{3+}$.

To further investigate the relation between the features of the TDoS of Li$_x$NiO$_2$ and the oxidation states of the Ni cations it is instructive to look at the Ni PDoS.

We can clearly distinguish the contributions of the differently oxidized Ni cations looking at the Ni PDoS shown in Fig. 13. In fully lithiated LNO, only Ni$^{2+}$ and Ni$^{4+}$ are present. The peaks A$_1$ and, at a much lesser extent, A$_1'$ in the occupied part and B$_2$ in the unoccupied one are due to Ni$^{2+}$ (see Fig. 4 (d)) while Ni$^{4+}$ (Fig. 4 (c)) is characterized by only two wide empty features A$_2$ (spin up) and B$_1$ (spin down) in roughly the same energy range (0.3 – 1.2) eV. As Li vacancies are created, the number of Ni$^{2+}$ cations decreases and that of Ni$^{3+}$ increases. We can see from Fig. 13 (b) that the Ni$^{3+}$ cations contribute to the A$_1'$ spin up occupied peak near the Fermi level and to the wide empty spin up feature A$_2$ at about 0.5 eV. Also the empty spin down split peaks B$_1'$, and B$_1''$ in the energy range from 1 eV to 2 eV are attributed to Ni$^{3+}$. These features can be compared with the Ni PDoS of the *C2/m* structure (Fig. 4 (e) and (f)), where only Ni$^{3+}$ cations are present. The B$_1'$ and B$_1''$ peaks widen and slightly shift with the delithiation but are recognizable up to $x = 0.25$ after which they start to disappear towards the full delithiation limit where only fully oxidized Ni$^{4+}$ cations are present. With progressive delithiation we see that the occupied spin up $e_g^*$ features change: the A$_1$ peak (Ni$^{2+}$) decreases while A$_1'$ (Ni$^{3+}$) increases. At Li concentration lower than $x = 0.50$ the A$_1'$ peak decreases and disappears. The PDoS of Ni$^{4+}$ (Fig. 12 (c)) shows wide empty spin up and spin down features extending from 0 to 1.5 eV similar to the TDoS of NiO$_2$ (Fig. 10 at $x = 0$). The peaks and features of the Ni PDoS change with the increasing of the delithiation, in particular the Ni$^{2+}$ and Ni$^{3+}$ PDoS at $x = 0.25$ undergo a considerable modification. One main reason is the shrinking of the cell volume (see Fig. 9) leading to the shortening of the Ni-O bonds and related increasing of the Ni-O interaction. For the antibonding occupied $e_g^*$ orbitals it means a shift towards higher energies. Furthermore, the presence of residual JT distorted orbitals in the shrinked volume leads to enlarged splittings of the bands. Indeed, the presence of Li vacancies introduces further differences in the strengths of the Ni-O bonds, which depend also on the different charges on the Oxygen ions, with the less charged oxygens being those bonded to less Li ions.

The estimated local magnetic moments on the Ni ions in L$_x$NiO$_2$, are listed in Table S1 (SI).

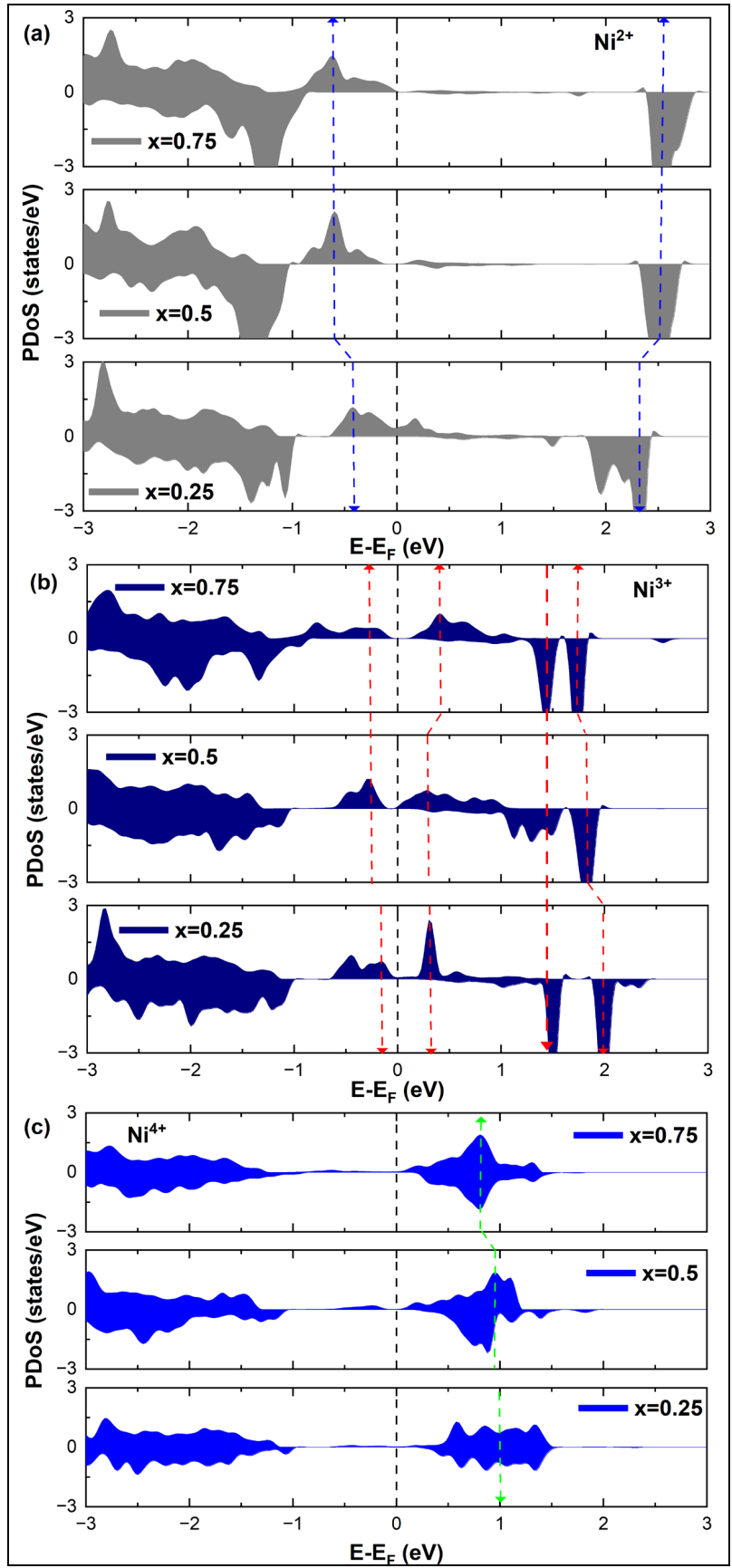

Fig. 13. {(a), (b), (c)} Projected DOS (PDOS) of the Ni-3d with different magnetic moments in $Li_xNiO_2$.

## 5. Mn for Ni substitution in $Li_xNiO_2$

**5a. Substitutional Mn Mixing Energies and Structural Properties**

We substituted 6, 12, and 18 Ni ions over the 48 transition metal sites in the template cell with Mn concentrations $y$ = 0.125, 0.25, and 0.374 of $Li_xNi_{(1-y)}Mn_yO_2$ all within the concentration range for which LNMO compounds have been synthesized in the α-$NaFeO_2$ layered structure [50, 51, 52].

The different Mn in-plane arrangements are reported in Fig. 14. Mn ions are arranged in the same way in all the transition metal layers of the O3 stacking with detailed information given in Table S4.

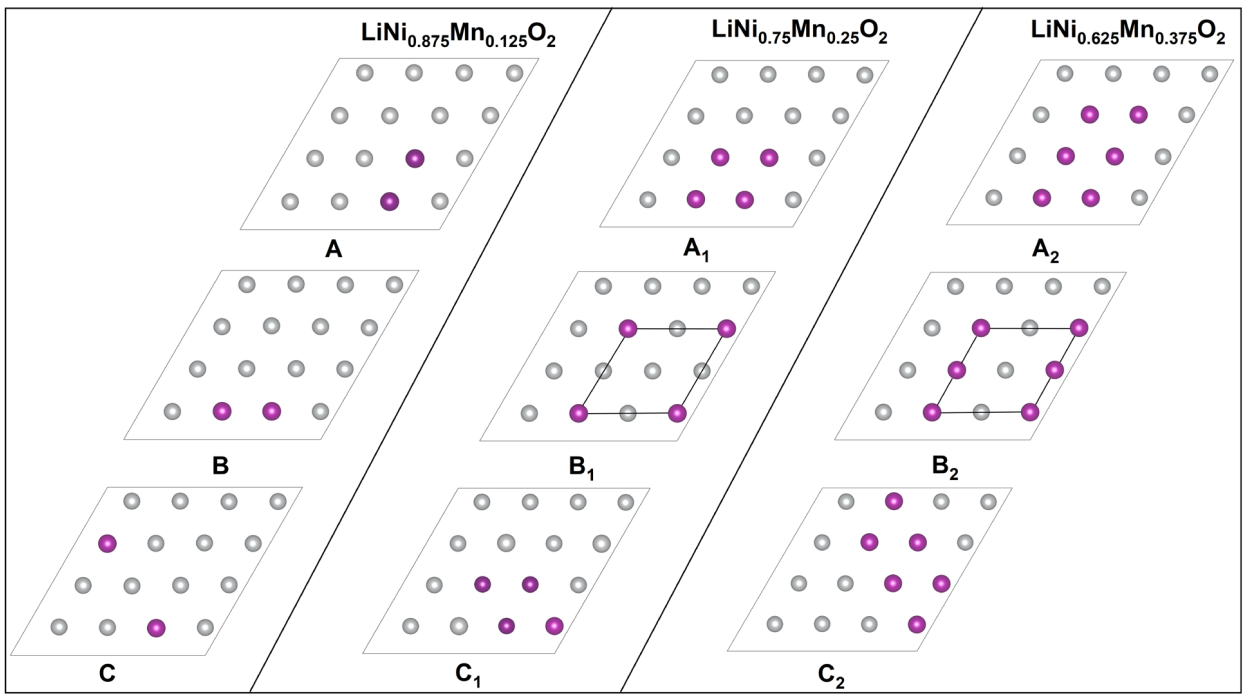

Fig.14 Configurations of Mn ions in the transition metal planes. The purple dots show the positions of the Mn ions.

A, B, C correspond to the 12.5% concentration of Mn atoms, $A_1$, $B_1$, $C_1$ to 25%, and $A_2$, $B_2$, $C_2$ to 37.5%. The lowest energy configurations for each Mn concentration in the fully lithiated case were found to be A, $B_1$ and $B_2$ and we use them to optimize the 192 atom supercell for different concentrations of Li vacancies obtaining the optimized lattice parameters reported in Table S5.

To compare with the results of the previous section we have considered the same delithiated patterns. In Fig. 15, we report the trend of the calculated mixing energies as for Eq. 2. It is clear that Mn substitution is particularly favorable at low concentrations of Li vacancies, while below $x$ = 0.25

(75% Li vacancies) the presence of Mn becomes less advantageous and for $x < 0.2$ the compounds with highest Mn concentrations are unstable against phase separation. These results are qualitatively in agreement with the trend of previous calculations which considered only the $x = 1$ and $x = 0$ cases [53]. This result was interpreted in terms of an attractive interaction between the Ni and Mn ions at full lithiation becoming a repulsive one at full delithiation. This result seems also apparently contrary to experimental findings where the Mn presence seem to have a beneficial effect just at high cathode charge voltages. The discrepancy is apparent since in literature the stabilization effect of Mn is attributed to the pillar effect, that is, to the fact that Mn at low concentrations would induce more $Ni^{2+}$ defects in the Li layer than pristine LNO and would favor $Li^+/Ni^{2+}$ mixing. It was shown [54] that the number of these $Ni^{2+}$ defects in the Li layer remains unchanged during charge and discharge cycles and their presence would avoid the collapse of the Li intercalation layers at high delithiation levels. Our calculations do not consider here the presence of such defects, thus, the trend shown in Fig. 15 refers to the case of ideal cathodes without Ni/Li defects.

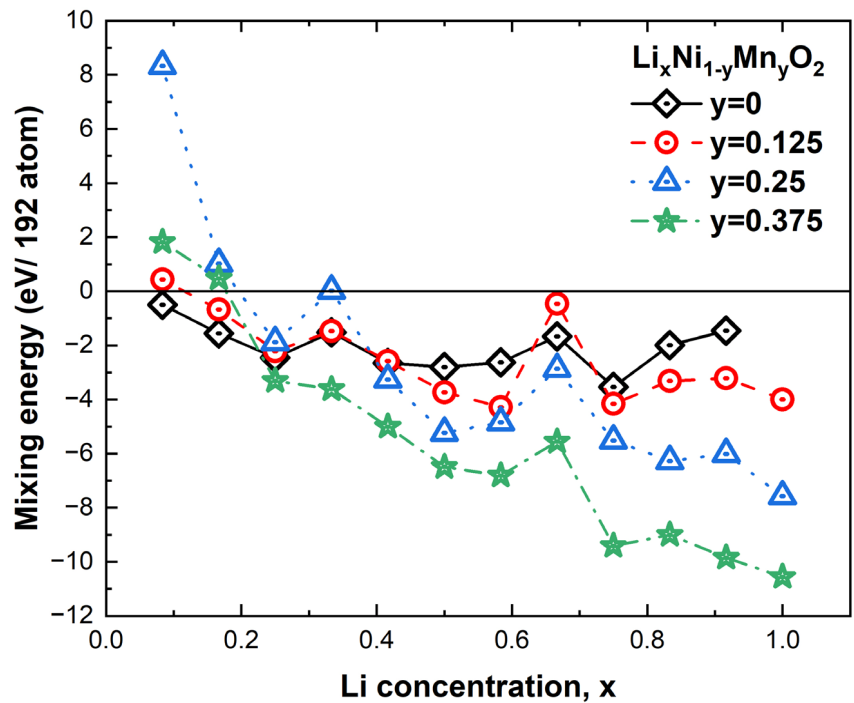

Fig 15. Mixing energies of $L_xNMO$ for different values of Li concentrations $x$.

The change of the lattice parameters of the supercell as a function of Li and Mn concentrations $x$ and $y$ are shown in Fig. 16. The $a$ and $c$ parameters follow the same trends with $x$ as $L_xNO$. Both $a$ and $c$ become longer with increasing Mn concentration. This is in agreement with the experimental results for $LiNi_{0.8-x}Co_{0.1}Mn_{0.1+x}O_2$, with $0.0 \leq x \leq 0.08$, where the lattice parameters, especially the $c$ lattice parameter, increases with increasing Mn content. The increase was said to be due to the substitution of $Ni^{3+}$ cations by $Mn^{4+}$ ions, which have a higher ionic radius, and the consequent reduction of a corresponding number of $Ni^{3+}$ ions into larger $Ni^{2+}$ (0.69 Å) to maintain charge neutrality [55, 56]. This interpretation stems again from the assumption that in pristine LNO all the Ni are in a $Ni^{3+}$ oxidation state. $Ni^{2+}$, or almost $Ni^{2+}$, are already present in pristine LNO in our case, thus, the reason for the larger parameters is due, as we will see below, to the partial substitution of $Ni^{4+}$ by the larger $Mn^{4+}$ ions and the increase of the total number of electrochemically active ($Ni^{2+}$ + $Ni^{3+}$) cations to which are associated larger octahedra (see below Table 4). Apart the very first value ($x = 1.0$) the $a$ parameter of LNO ($y = 0$) follows this trend with the shortest $a$ values while the behavior of $c$ does not. Indeed only for $y = 0.375$ the $c$ lattice parameter of $Li_xNi_{(1-y)}Mn_yO_2$ is longer than that of the system without Mn ($y = 0$). This non-linear trend of the $c$ parameter with $y$ is most likely due to the fact that the A, $B_1$, and $B_2$ Mn arrangements are inserted in lattices having for each $x$ different delithiation patterns. As we will see later the delithiation pattern and the Mn distribution pattern strongly affect each other in determining the electronic configuration of the systems. In this paper we have not searched for the lowest energy structures where the total energy is minimized for both $x$ and $y$ simultaneously, so we do not expect a very precise trend in the calculated parameters.

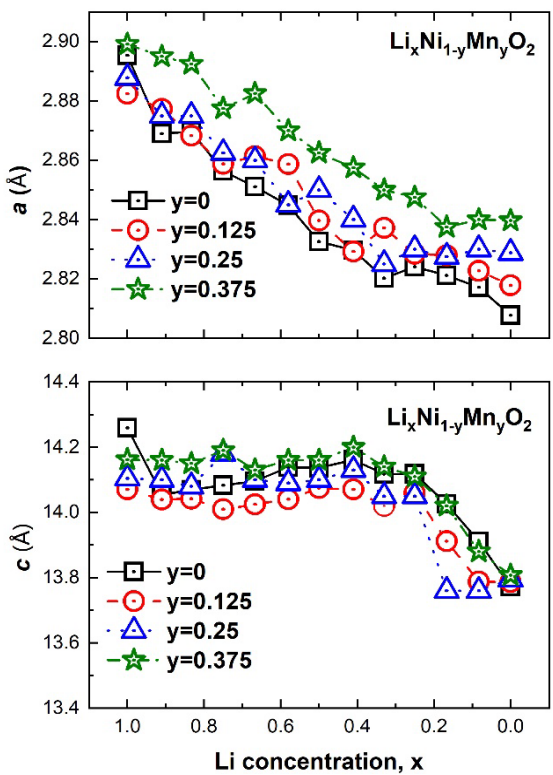

Fig. 16. Crystal lattice parameters versus Li concentration $x$ of $Li_xNi_{(1-y)}Mn_yO_2$ for the three Mn concentrations: $y$ 12.5%, 25% and 37.5%, respectively.

## 5b. Electronic and magnetic properties

First of all, looking at the estimated magnetic moments of the transition metals, Table S6, we have verified that, as is being generally assumed in the literature, the Mn ions are in the $Mn^{+4}$ inactive oxidation state, in agreement with other calculations [53] and most experiments [56].
Table 4 reports the evolution of the number of the differently oxidized transition metals as a function of the Li and Mn concentrations $x$ and $y$.

Table 4. Number of different Ni ions and Δ parameter in $Li_xNi_{1-y}Mn_yO_2$.

| $Li_xNi_{1-y}Mn_yO_2$ | | No. of oxidized Ni at Mn substitution | | | Jahn-Teller distortion parameter Δ in $Li_xNi_{1-y}Mn_yO_2$ | | | |
|---|---|---|---|---|---|---|---|---|
| $y$ | $x$ | $Ni^{2+}$ | $Ni^{3+}$ | $Ni^{4+}$ | $Ni^{2+}$ | $Ni^{3+}$ | $Ni^{4+}$ | $Mn^{4+}$ |
| 0 | 1 | 27 | 0 | 21 | 0.065 | - | 0.04 | - |
| 0.125 | 1 | 18 | 12 | 12 | 0.06 | 0.21 | 0.04 | 0.06 |
| | 0.75 | 10 | 16 (14 up, 2 down) | 16 | 0.1 | 0.21 | 0.03 | 0.05 |
| | 0.5 | 2 | 20 | 20 | 0.03 | 0.19 | 0.06 | 0.06 |
| | 0.25 | 0 | 12 | 30 | - | 0.19 | 0.05 | 0.05 |
| 0.25 | 1 | 12 | 24 | 0 | 0.02 | 0.22 | - | 0.01 |
| | 0.75 (Chain) | 2 | 32 (20 up, 12 down) | 2 | 0.07 | 0.25 | 0.06 | 0.05 |
| | 0.75 (Square) | 12 | 12 | 12 | 0.04 | 0.27 | 0.06 | 0.05 |
| | 0.5 | 2 | 20 (2 up, 18 down) | 14 | 0.1 | 0.23 | 0.06 | 0.05 |
| | 0.25 | 0 | 12 (1 up, 11 down) | 24 | - | 0.17 | 0.03 | 0.05 |
| 0.375 | 1 | 18 | 12 | 0 | 0.05 | 0.21 | - | 0.05 |
| | 0.75 | 9 | 18 | 3 | 0.07 | 0.21 | - | 0.08 |
| | 0.5 | 4 | 15 | 11 | 0.1 | 0.25 | 0.03 | 0.09 |
| | 0.25 | 2 | 8 | 20 | 0.06 | 0.18 | 0.06 | 0.09 |

From Table 4, we can obtain the following information:

(i) The number of the active electrons $N$ in the $e_g^*$ orbitals are determined by the Li concentration. Indeed, if we assign two electrons to $Ni^{2+}$ and one to $Ni^{3+}$ we see that for $x = 1$ we have about $N = 48$ electrons in $e_g^*$, for $x = 0.75$ $N = 36$, for $x = 0.50$ $N = 24$, for $x = 0.25$ $N = 12$, for all $y$. This result means that the missing electrons due to the delithiation are again taken mainly from the Ni cations.

(ii) The number of $Ni^{4+}$ cations diminish while the $Mn^{4+}$ cations increase. Their sum remains approximately the same and increase with the increase of the delithiation.

(iii) Comparing similarly delithiated structures, at $x = 0.75$ and 1, with and without Mn/Li substitution, we see an increase of the electrochemically active Ni ions, in particular $Ni^{3+}$.

(iv) Differently from the results of Table 3 above, for $y \neq 0$, the particular delithiation arrangement matters in the redistribution of the electrons and the number of $Ni^{2+}$, $Ni^{3+}$, and $Ni^{4+}$ cations. This is shown by the two entries corresponding to $x = 0.75$ and $y = 0.25$ where only the in-plane arrangement of Li vacancies is different, the chain and the square patterns examined above. This means that the arrangement of Mn ions and the arrangement of Li vacancies interact. The square Li vacancy arrangement together with the square arrangement of Mn ions produces the most stable of the structures studied for this same $x$ and $y$ concentration.

(v) The relation between $Mn^{4+}$ ions and Li vacancies is also seen in a few antiferromagnetic couplings arising in the case of chain delithiation and low Mn concentrations. In the case of chain delithiation and low Mn concentration the structural optimization produces a number of singly occupied spin down $e_g^*$ orbitals on Ni ions previously being $Ni^{4+}$ ions in the absence of Mn. The antiferromagnetic coupling between the Mn and Ni ions are in this case mediate by the less charged $O''$ oxygen atoms, first neighbors to two Li vacancies, present only in the chain delithiation model. The $Ni^{3+}$ cations with occupied spin down $e_g^*$ orbitals are, in absence of Mn ions, $Ni^{4+}$ ions arranged in the chains bonded to four $O''$ ions (two below and two above) shown in Fig. 12 (a). In this case electron transfer occurs from and to the $e_g^*$ orbitals through the $p\sigma$ orbitals of $O''$ [36, 57]. However, for Li concentrations as high as $x = 0.75$ the chain delithiation model is less stable than the square model, where this superexchange effect is not taking place. On the other side, at high levels of delithiation, the $O''$ oxygen ions could play an important role in mediating the interaction between Ni and Mn which is supposed to become repulsive in this limit.

In order to analyze the global effect on the electronic structure of Mn doping, we have calculated the TDoS and local magnetic moments. In Fig. 16 we report the modification of the TDoS with the degree of delithiation. For the occupied $e_g^*$ orbitals near the Fermi Level we see again progressive diminishing of the A peaks. Between the $B_1$ and $B_2$ empty peaks a number of new structures emerge, some due to the emergence of the $Ni^{3+}$ ions as seen above and others due to the $Mn^{+4}$ ions. It is shown in Fig. 17 that the Mn $e_g^*$ orbitals (hybridized Mn $d$ -O $p$ ) are empty and fall above 1 eV, with a spin up feature between 1 eV and 2 eV (in correspondence with the spin down peaks due to $N^{3+}$) and two negative features between 2 eV and 3.5 eV (the first one in correspondence with the spin down structures of $N^{2+}$). These features become larger and merge with the increase of Mn concentration. There are not occupied Mn $e_g^*$ orbitals in agreement with the assigning of the $Mn^{4+}$ oxidation state to the Mn ions. Around the Fermi Level the active electronic $e_g^*$ orbitals are still provided by the hybridized Ni $d$ – O $p$ obitals where the Ni ions are $Ni^{2+}$ and $Ni^{3+}$. The inspection of the Ni ions PDoS around the Fermi Level shows similar features to those already reported in Fig. 13. These results confirm that Mn in $Li_xNi_{(1-y)}Mn_yO_2$ is electrochemically inactive and its only role is to stabilize the structure. Indeed it was found that the disorder Ni/Li tends to decrease during cycling and not to increase which is one reason of the larger stability provided by Mn ions where the fixed number of $Ni^{2+}$ in the Li layer sustains the structure from collapse at high charge levels [54, 58, 59].

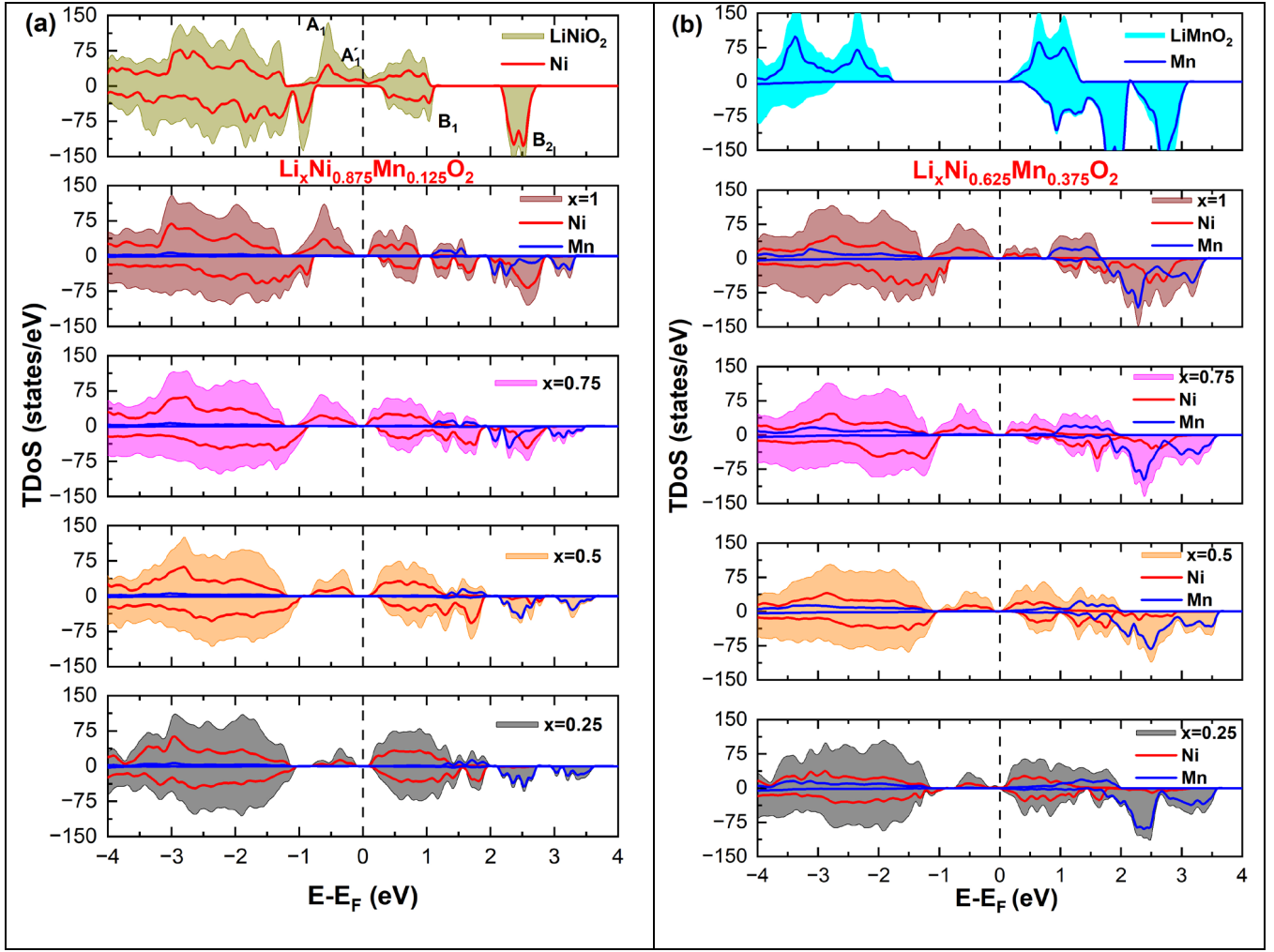

Fig. 17 (a, and b) Total Density of states (TDoS) of Mn substituted LNO as a function of Li concentration $x$ for $y = 0.125$ and $0.375$. The TDoS of LiMnO$_2$ refers to ferromagnetic ordering for comparison reasons. Antiferromagnetic ordering between Mn ions is indeed more stable.

It is generally thought that $Mn^{4+}$ are responsible for the generation in LNO of $Ni^{2+}$ cations since, for charge neutrality in layered LiNiO$_2$ and LiMnO$_2$ the transition metal cations have to be in the $Ni^{3+}$ and $Mn^{3+}$, respectively. Thus, the presence of $Mn^{4+}$ ions should be balanced with the simultaneous reduction of $Ni^{3+}$ to $Ni^{2+}$ to compensate for the change of $Mn^{3+}$ to $Mn^{4+}$. This observation has arisen questions about how to explain the fact that the discharge potential measured in layered LiNiO$_2$ and LiNi$_{0.5}$Mn$_{0.5}$O$_2$ in the hexagonal phase are practically identical [60, 61] given that different redox reactions are expected in LiNiO$_2$, where only $Ni^{3+}/Ni^{4+}$ is supposed to be active, and in LiNi$_{0.5}$Mn$_{0.5}$O$_2$, where both $Ni^{2+}/Ni^{3+}$ and $Ni^{3+}/Ni^{4+}$ should occur [53]. The implication of only $Ni^{3+}$ ions being present in LNO is due to the fact that the calculations were performed using small supercells for both rhombohedral and monoclinic symmetries and these calculations produce invariably only $Ni^{3+}$ ions, as we have seen above. Also the few percent presence of $Ni^{2+}$ defects in the Li layer of LNO cannot explain the discrepancy but the charge disproportionation we have found using the supercell can, since the number of $Ni^{2+}$ are similar in LNO and in $LiNi_{0.5}Mn_{0.5}O_2$ and the charging is going to create similar numbers of $Ni^{3+}$ and $Ni^{4+}$ ions in both systems.

## 6. Summary and Conclusions

In summary this paper deals with the layered oxide LNO and LMNO materials proposed for Ni rich cathodes. We have first replicated 16 times the primitive cell of the $R\bar{3}m$ space group, the structure of LNO observed experimentally by XRD, and relaxed both cell parameters and internal positions. The supercell relaxation leads to a large decreasing of the total energy accompanied by a charge and size disproportionation of the $NiO_6$ octahedra. The charge disproportionation has a large effect on the Ni oxidation states. While in the smaller primitive unit cell all Ni ions are $Ni^{3+}$, in the supercell there are only $Ni^{2+}$ and $Ni^{4+}$ ions. The $Ni^{2+}$ and $Ni^{4+}$ ions arrangement forms an ordered superstructure in the *ab* layers and the superstructure still belongs to the observed hexagonal symmetry. The Ni-O bond lengths form a distribution, as found experimentally, around two average bond lengths whose values are also in agreement with the experiment, and contrary with the behavior expected for a monoclinic crystal structure. The gain in energy obtained in the atomic relaxation of the supercell compared to that of the primitive cell is much larger than the one obtained with an analogous supercell of monoclinic symmetry, although the monoclinic structure remains still more stable. We conclude that the choice of the unit cell is important for this class of materials and an even larger supercell, or perhaps the inclusion of Ni defects in the Li layer (always present in the real samples) may favor the observed hexagonal phase over the monoclinic one at full lithiation. We have found that optimizing larger unit cells leads to emergence of new structural and electronic behaviors which remain hidden if using the primitive or smaller cells.

Then, we have used the calculated superstructure as a template for the study of the delithiation in the hexagonal LNO structure. We have calculated the energies of a number of structures for different concentrations of Li and studied the evolution of the structural parameters and the lectronic structure with the number of Li vacancies. The changes of the *a* and *c* lattice parameters are in agreement with previous calculations, and with the experiments, at least for the Li concentration for which LNO is in the hexagonal structure. We have found that delithiation leads to changes in the number of Ni ions, in particular $Ni^{2+}$ become $Ni^{3+}$ and $Ni^{4+}$, the dominant Ni ion at high battery charging levels. $Ni^{3+}$, and its related JT octahedral distortions, become the dominant ions at the Li concentrations of LNO in the monoclinic phase.

Finally, we have analyzed the changes in the properties of the delithiated structures when the 12.5%, 25% and 37.5% of Ni ions are substituted with Mn ions. We have found that the way Mn affects the lattice parameters and electronic structure depends on the arrangement of both the Mn ions in the transition metal layers and of the Li vacancies in the Li layers. However, in all cases, the Mn ions are always in the $Mn^{4+}$ oxidation state and do not participate actively in the redox reactions determining the battery charge and discharge potentials, since the important electronic states around the Fermi level involved in the reactions belong only to the $Ni^{2+}$ and $Ni^{3+}$ ions. The presence of a large number of $Ni^{2+}$ ions in pristine LNO can explain the similarity between the LNO and the $LiNi_{05}Mn_{0.5}O_2$ discharge potentials found experimentally since both $Ni^{2+}/Ni^{3+}$ and $Ni^{3+}/Ni^{4+}$ redox pairs can be present in the initial stages of charging in both systems.

## Acknowledgments

This research was developed under the framework of the BAT4EVER project that has received funding from the European Union's Horizon 2020 research and innovation program under Grant Agreement No 957225.  The authors would also like to acknowledge the CINECA HPC facility for the approved ISCRA B (IscrB_CATCH22 and IscrB_LION-CAT). R.M. acknowledges the PNRR MUR project ECS_00000033_ECOSISTER. R.M. acknowledges the MOST e Sustainable Mobility Center  funded by the European Union Next-Generation EU (PIANO NAZIONALE DI RIPRESA E RESILIENZA (PNRR) e MISSIONE 4 COMPONENTE 2, INVESTIMENTO 1.4 e D.D. 1033 June 17, 2022, CN00000023). This manuscript reflects only the authors' views and opinions, neither the European Union nor the European Commission can be considered responsible for them.

## Author's Contributions

**Saleem Yousuf**  Performed calculations. Contribution in writing the paper. Made Figures and Tables. Interpretation of simulation results.

**Md Maruf Mridha**   Performed Calculations

**Rita Magri**  Conceptualization. Validation calculations. Interpretation of simulation results. Writing of paper. Supervision. Acquisition of funding.

## Data availability

The raw data required to reproduce these findings are available upon request.